\documentclass[twocolumn]{aastex631}

\newcommand{\name}{COSMOS2020-635829}
\newcommand{\oiifull}{[O\textsc{ii}]$\mathrm{\lambda \lambda 3726-3729}$}
\newcommand{\oii}{[\textsc{Oii}]}

\usepackage{threeparttable}
\usepackage{booktabs}
\usepackage{amsmath}
\usepackage{color}
\usepackage{soul}
\usepackage{bm}
\usepackage{placeins}

\begin{document}

\title{JWST Reveals a Candidate Jellyfish Galaxy at $\bm{z=1.156}$}

\correspondingauthor{Ian D. Roberts}
\email{ianr@uwaterloo.ca}

\author[0000-0002-0692-0911]{Ian D. Roberts}
\affiliation{Department of Physics \& Astronomy, University of Waterloo, Waterloo, ON N2L 3G1, Canada}
\affiliation{Waterloo Centre for Astrophysics, University of Waterloo, 200 University Ave W, Waterloo, ON N2L 3G1, Canada}

\author[0000-0003-4849-9536]{Michael L. Balogh}
\affiliation{Department of Physics \& Astronomy, University of Waterloo, Waterloo, ON N2L 3G1, Canada}
\affiliation{Waterloo Centre for Astrophysics, University of Waterloo, 200 University Ave W, Waterloo, ON N2L 3G1, Canada}

\author[0000-0003-0780-9526]{Visal Sok}
\affiliation{Department of Physics and Astronomy, York University, 4700 Keele Street, Toronto, ON M3J 1P3, Canada}

\author[0000-0002-9330-9108]{Adam Muzzin}
\affiliation{Department of Physics and Astronomy, York University, 4700 Keele Street, Toronto, ON M3J 1P3, Canada}

\author[0000-0002-1437-3786]{Michael J. Hudson}
\affiliation{Department of Physics \& Astronomy, University of Waterloo, Waterloo, ON N2L 3G1, Canada}
\affiliation{Waterloo Centre for Astrophysics, University of Waterloo, 200 University Ave W, Waterloo, ON N2L 3G1, Canada}
\affiliation{Perimeter Institute for Theoretical Physics, 31 Caroline St N, Waterloo, ON N2L 2Y5, Canada}

\author[0000-0002-9655-1063]{Pascale Jablonka}
\affiliation{Laboratoire d'astrophysique, {\'E}cole Polytechnique F{\'e}d{\'e}rale de Lausanne (EPFL), CH-1290 Sauverny, Switzerland}



\begin{abstract}
We report the discovery of \name{} as a candidate jellyfish galaxy undergoing ram pressure stripping in a (proto)cluster at $z > 1$. High-resolution imaging from the \textit{James Webb Space Telescope} reveals a symmetric stellar disk coupled to a unilateral tail of star-forming knots to the south. Using Gemini GMOS IFU observations, we show that these extra-planar continuum sources are embedded within an ionized gas tail that is kinematically connected to the disk of \name{}. If confirmed, this represents the highest-redshift discovery of a ram pressure stripped ionized gas tail. The tail sources are characterized by extremely young stellar populations ($\lesssim 100\,\mathrm{Myr}$), have stellar masses of ${\sim}10^8\,\mathrm{M_\odot}$, and star formation rates of $0.1\text{--}1\,\mathrm{M_\odot\,yr^{-1}}$. This work shows that ram pressure stripping can potentially perturb group and cluster galaxies at $z > 1$ and may contribute to environmental quenching even near Cosmic Noon. 
\end{abstract}

\keywords{}


\section{Introduction} \label{sec:intro}

For approximately 50 years it has been known, at least to some degree, that the properties of galaxies in overdense environments like galaxy groups and clusters differ systematically from those relatively isolated in the field. This dichotomy is reflected in a number of galaxy properties, but among the most studied is the overabundance of red, quiescent satellite galaxies in dense environments \citep[e.g.][]{dressler1980,postman2005,blanton2009,peng2010,wetzel2012,wilman2012,haines2015,roberts2017}. This `environmental quenching' of galaxies is strongest in the most-massive galaxy clusters with the fraction of quiescent galaxies approaching the field value for the lowest-mass galaxy groups \citep[e.g.][]{kimm2009,wetzel2012}.
\par
The origins of this quenching has been attributed to a suite of physical mechanisms that are unique to dense environments (see \citealt{cortese2021} for a recent review), the most commonly invoked being: the cessation of cold-gas accretion onto satellite galaxies (often referred to as `starvation' or `strangulation'), either due to the high virial temperature of the group or cluster halo or due to the circum-galactic medium being stripped off of the galaxy \citep[e.g.][]{larson1980,balogh1999,peng2015}; or the direct stripping of cold-gas from galaxies, either due to ram pressure stripping (RPS, e.g.\ \citealt{gunn1972,gavazzi1978,gavazzi1987_a1367}) or from tidal interactions with other group/cluster galaxies that can either directly strip material or tidally stir/heat gas making it more easily removed by ram pressure \citep[e.g.][]{boselli2006,mayer2006}. It is likely that the effectiveness of these various environmental perturbations depends on the mass of the host halo. For example, ram pressure strength scales with the density of, and velocity relative to, the ICM as $\rho_\mathrm{ICM} v^2$, and thus will be strongest in massive galaxy clusters. Conversely, tidal interactions and even galaxy mergers will be more common in lower mass groups where the typical orbital velocities are closer to the escape velocity for two interacting galaxies.
\par
In the low-redshift Universe, a comprehensive understanding of the relative balance between different quenching mechanisms as a function of host environment is still elusive -- but progress has been made. It is now clear that RPS plays an important role in quenching galaxies in galaxy clusters. This has been inferred through the need for short quenching times \citep[e.g.][]{quilis2000, muzzin2014}, observed signatures of outside-in quenching \citep[e.g.][]{cortese2012,schaefer2017,finn2018,owers2019,schaefer2019,morgan2024,broderick2025}, as well as the identification of so-called `jellyfish' galaxies with one-sided tails of stripped interstellar medium that extend well beyond the main galaxy disk \citep[e.g.][]{gavazzi1987_a1367,gavazzi2001,chung2007,poggianti2017,boselli2018,roberts2021_LOFARclust,roberts2022_perseus,ignesti2023_a2255}. In lower mass galaxy groups, the picture is less clear. RPS certainly does occur in such halos \citep[e.g.][]{bureau2002,mcconnachie2007,rasmussen2006,hu2024,roberts2024_2276,finn2025}, but the frequency is lower \citep{roberts2021_LOFARgrp}. Even in the most-massive clusters, it is unlikely that ram pressure is responsible for stripping all (or even most) of the molecular gas in satellite galaxies \citep[e.g.][]{zabel2022,watts2023,brown2023}. Instead, quenching likely operates in conjunction with gas depletion (which is then not replenished) from star formation, both prior to galaxies reaching the inner regions of groups and clusters where RPS is efficient, and after gas disks are truncated by ram pressure in order to `finish off' quenching. This has recently been framed as a `slow-then-rapid' quenching paradigm where this interplay between starvation/strangulation and RPS drives the quenched fractions in low-redshift clusters \citep{wetzel2013,maier2019,vanderburg2018,roberts2019,manuwal2023,morgan2024,broderick2025}.
\par
At higher redshift ($z \sim 1\text{--}2$), far less is known. It has been shown that cluster red sequences in excess of the field population are in place in dense environments at these redshifts \citep[e.g.][]{muzzin2012,cerulo2016,old2020}. This, in and of itself, implies that relatively short quenching timescales are needed given the limited time available for group/cluster formation and subsequent environmental quenching at high-$z$. Direct evidence for RPS at high-redshift is limited due to both the required sensitivity to detect the typically faint morphological signatures of RPS and the required angular resolution for spatially resolving them. Of the examples in literature, \citet{noble2019} present marginally resolved molecular gas extensions in $z \sim 1.6$ cluster galaxies that may represent ram-pressure stripped tails. More recently, \citet{xu2025} report molecular gas observations of galaxies in a $z=2.51$ protocluster, at least three of which show some compelling evidence for one-sided tails, potentially driven by ram pressure.
\par
The bulk of galaxies on the cluster red sequence at $z \sim 0$ ceased substantial star formation at $z \gtrsim 1\text{--}2$ \citep[e.g.][]{kodama1997,depropris1999,mancone2010}. Thus a key avenue for understanding the quenched population in the local Universe is to observe environmental quenching mechanisms, such as RPS, in action at high-$z$. In the local Universe, ionized gas tails are among the most frequently identified signatures of RPS \citep[e.g.][]{yagi2010,poggianti2017,boselli2018}, but to-date no examples of ram-pressure stripped ionized gas tails have been identified beyond $z=0.7$ \citep{boselli2019}. Identifying analogous galaxies at high-redshift (to the extent that they exist) must be a fruitful path for advancing our knowledge of environmental quenching. Currently the number of known, unquestionable examples of RPS at $z > 1$ is likely zero, and the number of jellyfish galaxy candidates in the present literature is almost certainly ${<}10$. Any new identification of such objects is thus an important addition.
\par
In this work we present \name{} ($\alpha=149.81480,\;\delta=2.02811$), a new candidate for a satellite galaxy at $z = 1.156$ undergoing active RPS. \name{} is associated with a cluster-mass, X-ray detected overdensity and shows a one-sided collection of blue, extra-planar knots that are co-spatial with an ionized gas tail. This is among the strongest candidates for a jellyfish galaxy at $z>1$, and if said interpretation is correct, the current highest redshift example of a ram-pressure stripped ionized gas tail as well extra-planar star formation.
\par
The outline of the paper is as follows. In Sect.~\ref{sec:data} we introduce the data products used in this work, including broadband imaging (JWST, HST, Subaru) as well as spatially resolved spectroscopy (Gemini). In Sect.~\ref{sec:JFcand} we provide an outline of \name{}'s status as a jellyfish galaxy candidate based on rest-frame optical and near-IR imaging. In Sect.~\ref{sec:gas_tail} we present the detection of an ionized gas tail to the south of \name{} that is co-spatial with the observed rest-frame optical asymmetries. In Sect.~\ref{sec:tail_sf} we present an analysis of the ongoing star formation in the extra-planar sources, presumed to be fueled by ram-pressure stripped gas. Finally, in Sects.~\ref{sec:discussion} and \ref{sec:conclusions} we discuss the key results from this work as well as make concluding remarks. Throughout, we assume a flat $\mathrm{\Lambda CDM}$ cosmology with $\Omega_M=0.3$, $\Omega_\Lambda=0.7$, and $H_0 = 70\,\mathrm{km\,s^{-1}\,Mpc^{-1}}$.

\section{Data} \label{sec:data}

\begin{figure*}[!ht]
    \centering
    \includegraphics[width=\textwidth]{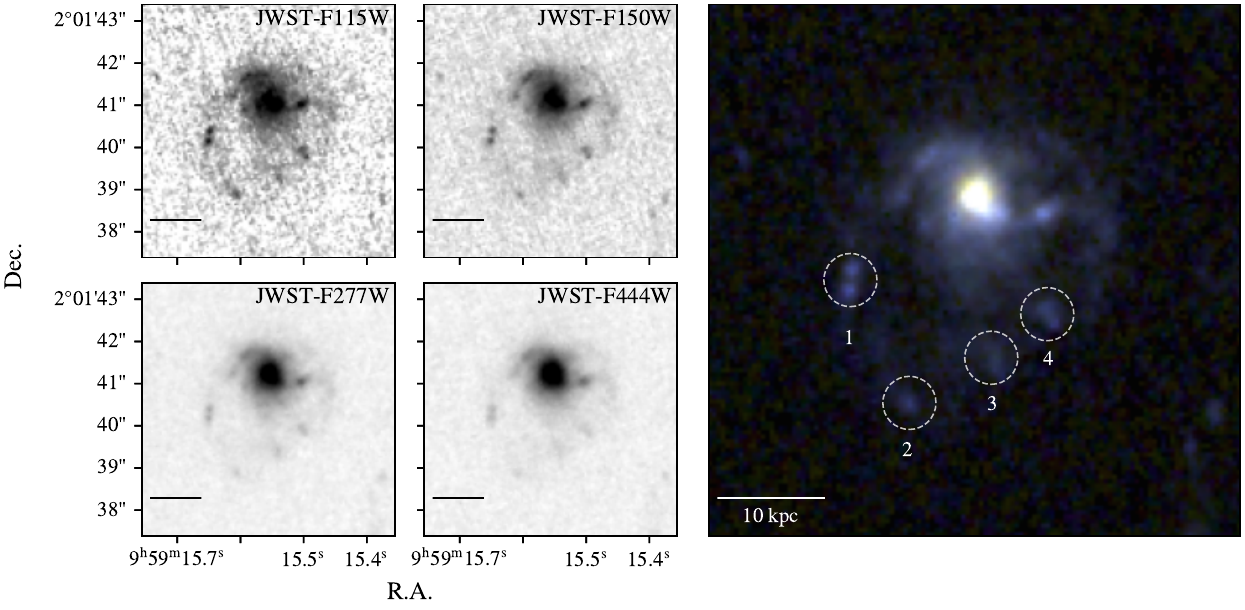}
    \caption{Thumbnail images of \name{} for the 4 JWST filters used in this work. The RGB image on the right is a combination of the JWST F444W (red channel), the F277W (green channel), and F115W+F150W (blue channel). The dashed circles mark the four extra-planar sources that are identified in the tail of \name{}.}
    \label{fig:panels}
\end{figure*}

\subsection{Broadband Imaging} \label{sec:data:imaging}

Our primary imaging source for this work is publicly available, multi-band \textit{James Webb Space Telescope} (JWST) imaging of the COSMOS field from the COSMOS-Web Survey \citep{casey2023,franco2025}. This is from COSMOS-Web Data Release 0.5 which includes ${\sim}0.3\,\mathrm{deg}^2$ of the COSMOS field in the NIRCam F115W, F150W, F277W, and F444W filters. A far smaller area is also covered by the MIRI F770W filter, though this does not cover \name{}. This JWST imaging was used to first identify \name{} as a jellyfish galaxy candidate (see Sect.~\ref{sec:JFcand}, Roberts et al. in prep). We also make use of \textit{Hubble Space Telescope} (HST) F814W imaging \citep{koekemoer2011} and deconvolved Subaru Suprime-Cam B-, V-, $r$-, and $z$-band imaging \citep{sok2022} in Sect.~\ref{sec:tail_sf} for spectral energy distribution (SED) fitting. Finite-resolution deconvolution is applied to the deep Suprime-Cam imaging of the COSMOS field to achieve an angular resolution of $0.3\arcsec$, this procedure is described in detail by \citet{sok2022}. The JWST and HST data described above may be obtained from the MAST archive at \dataset[doi:10.17909/ph8h-qf05]{https://dx.doi.org/10.17909/ph8h-qf05} and \dataset[doi:10.17909/T9XW2Q]{https://dx.doi.org/10.17909/T9XW2Q}.

\subsection{GMOS Spectroscopy} \label{sec:data:gmos}

We observed \name{} with the GMOS IFU in the 2025A semester (PI: Roberts, program code: GN-2025A-FT-205). We used the single-slit GMOS IFU with a central wavelength of $8900\,\mathrm{\AA}$ (with $\pm 100\,\mathrm{\AA}$ spectral dithers) and the R400 grating. This gives a field-of-view of ${\sim}3.5\arcsec \times 5\arcsec$ and a spectral resolution of $R=3000$ ($\sigma_\mathrm{inst} \simeq 43\,\mathrm{km\,s^{-1}}$). Our main target is the redshifted \oiifull{} emission line doublet, which is among the brightest observable emission lines in the rest-frame optical spectrum. The $\mathrm{H\alpha}$ line is not observable from the ground for \name{} as it falls between the J and H atmospheric windows.  We observed \name{} for a total integration time of $5\,\mathrm{h}$ which was split over three nights (2025-03-02, 2025-04-21, 2025-05-26) in observing blocks of $5 \times 1200\,\mathrm{s}$, each bracketed by flat-field and CuAr arc observations. The standard star EG131 was also observed for spectrophotometric calibration. For the three nights the raw image-quality during our observations was 70th percentile, 70th percentile, 20th percentile, respectively, and the average airmass was 1.2, 1.5, 1.5.
\par
Data reduction was done using the Gemini IRAF software package following the \href{https://gmos-ifu-1-data-reduction-tutorial-gemini-iraf.readthedocs.io/en/latest/index.html}{GMOS IFU-1 Data Reduction Tutorial}. The standard reduction steps were completed, including bias and overscan correction, flat fielding, scattered light removal, cosmic-ray rejection, and sky subtraction. We use our standard star observation for flux calibration, and we futher refine our flux scale using the calibrated DESI DR1 spectrum of \name{}. A calibrated data cube was produced for each $1200\,\mathrm{s}$ science exposure, with pixel sizes of $0.1\arcsec$ in the spatial directions and $0.76\,\mathrm{\AA}$ in the spectral direction. The cubes were then reprojected onto a common wavelength grid and coadded with inverse-variance weighting to produce our final science cube. In the vicinity of the \oii{} doublet, we achieve a $1\sigma$ sensitivity of ${\lesssim}10^{-18}\,\mathrm{erg\,s^{-1}\,cm^{-2}\,\AA^{-1}\,arcsec^{-1}}$.

\section{\name{}: A Jellyfish Galaxy Candidate at $z > 1$} \label{sec:JFcand}

\name{} is a an intermediate mass ($M_\mathrm{star} \sim 10^{10}\,\mathrm{M_\odot}$), strongly star-forming ($\mathrm{SFR} \sim 100\,\mathrm{M_\odot\,yr^{-1}}$) galaxy at a redshift of $z_\mathrm{spec} = 1.1560$ \citep{weaver2022,desi2023}. It was initially flagged as a jellyfish galaxy candidate by Roberts et al. (in prep) based on high-resolution rest-frame optical and near-IR imaging from the JWST COSMOS-Web survey \citep{casey2023}. This selection of jellyfish galaxy candidates at $z \gtrsim 1$ will be outlined in detail by Roberts et al. (in prep), but in brief, galaxies were selected according to one or more of the following morphological features.
\begin{enumerate}
    \item A one-sided extension in the stellar distribution, which is interpreted as being in the direction of a potential ram pressure stripped tail. This extension may be associated with unwinding spiral arms \citep[e.g.][]{bellhouse2021}, but it does not need to be.

    \item A one-sided collection of knots that are detached from the main stellar distribution of the galaxy. We require that these knots are similar in color to at least some part of the main galaxy, in order to reduce the odds of background sources in projection.

    \item Evidence for enhanced star formation on the proposed leading edge. We require this to be in conjunction with a proposed tail direction from one of the above features. This means that a galaxy that shows apparently asymmetric star formation within its disk, but no sign of a one-sided extension, will not be classified as a jellyfish galaxy candidate. Signs of enhanced star formation on the leading edge may be used to classify a galaxy as a jellyfish galaxy candidate with only marginal evidence for a `tail' extension, if the orientation of the potential enhanced star formation and tail direction are mutually consistent.
\end{enumerate}
\noindent
This is analogous to the visual selections of ram pressure candidates that has become relatively common, and quite successful, at $z \sim 0$ \citep{ebeling2014,poggianti2016,roberts2020,durret2021,roberts2022_UNIONS,crossett2025}. In this framework the direction of the one-sided extension is interpreted as the direction of a \textit{potential} ram pressure tail. We stress that these visual selections will by no means identify a pure sample of galaxies undergoing RPS, as there are other physical processes (e.g.\ tidal effects) that can perturb galaxies in groups and clusters, but through follow-up observations and analyses we can hope to confirm some fraction of these candidates.
\par
In Fig.~\ref{fig:panels} we show single-band images of \name{} in the four COSMOS-Web filters as well as a color-composite. The morphology of \name{} is given by a symmetric main galaxy component coupled to compact, relatively blue, knots distributed to the south of the galaxy. It is these knots for which \name{} was selected as a ram pressure candidate under the assumption that they may correspond to sites of extra-planar star formation within a ram pressure stripped gaseous tail (i.e.\ criteria \#1 and \#2). While at first glance these knots appear largely detached from the main galaxy disk, close inspection (particularly of the F115W and F150W imaging) shows that they are connected to the main disk through faint emission that appears to extend from spiral arms in the main galaxy. The F115W (rest-frame NUV) is also brighter on the leading half than the trailing half of \name{}, and thus possibly also satisfying criterion \#3, though this was not a feature on which \name{} was initially selected. The four tail sources are highlighted in the color composite in Fig.~\ref{fig:panels} and will be the subject of further investigation in Sect.~\ref{sec:tail_sf}.

\begin{figure}[!ht]
    \centering
    \includegraphics[width=\columnwidth]{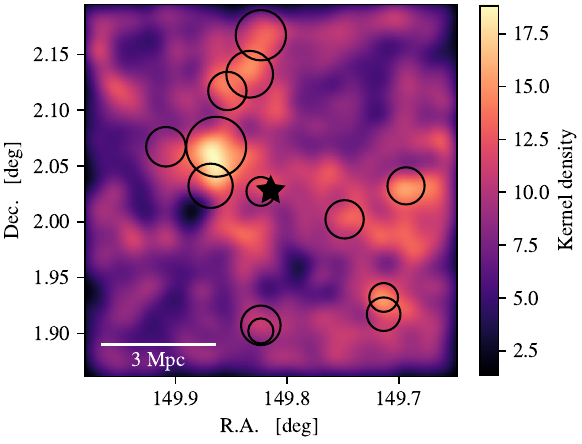}
    \caption{Large scale structure around COSMOS2020-635829. The large star marks the position of \name{}. The colormap shows a 2D kernel density estimate of all galaxies from the COSMOS2020 catalog within 10 arcminutes of \name{} and with $1.0 < z_\mathrm{phot} < 1.3$. Open circles correspond to groups/clusters from AMICO-COSMOS catalog within 10 arcminutes of \name{} and with $1.0 < z_\mathrm{phot} < 1.3$.}
    \label{fig:lss}
\end{figure}

The ram pressure candidates from Roberts et al. (in prep) are selected without knowledge of the environments within which the galaxies reside, but it is found post classification that the identified ram pressure candidates are preferentially associated with overdense environments. \name{} is no exception. In Fig.~\ref{fig:lss} we show the 2D galaxy density distribution for galaxies within $10\arcmin$ of \name{} and with $1.0 < z_\mathrm{phot} < 1.3$. \name{} is associated with a complex and extended overdensity. The galaxy is spatially coincident with a ${\sim}2-3 \times 10^{13}\,\mathrm{M_\odot}$ halo at $z_\mathrm{phot} \sim 1.23$, as identified by the AMICO cluster finding algorithm \citep{toni2024}. This halo also appears at the outskirts of a strong galaxy overdensity centered on two halos in the AMICO-COSMOS catalog with photo-$z$'s of 1.15 and 1.29, each X-ray detected at ${\sim}4\,\sigma$, and with a combined halo mass of ${\sim}10^{14}\,\mathrm{M_\odot}$.
\par
\begin{figure}
    \centering
    \includegraphics[width=\columnwidth]{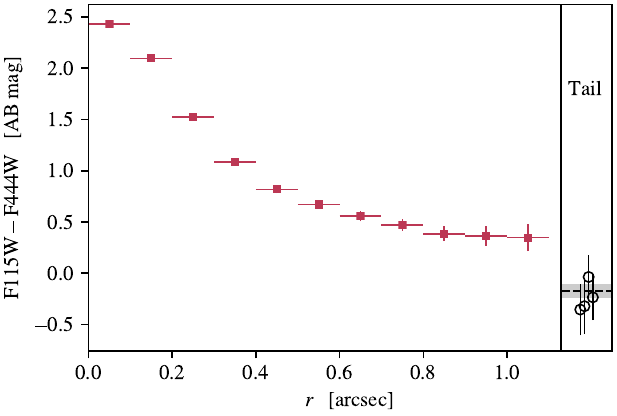}
    \caption{Color profile for the main disk of \name{} measured in circular annuli from the galaxy center to the edge of the disk. Vertical error bars show the $1\sigma$ statistical uncertainties on the measured color, horizontal error bars show the annuli widths. The far right of the figure shows the color in the tail region, both integrated over the full tail (dashed line) as well as measured individually for the tail sources highlighted in Fig.~\ref{fig:panels}.}
    \label{fig:gal_tail_col}
\end{figure}

The closest galaxy companion to \name{} (with $1.0 < z_\mathrm{phot} < 1.3$) has a projected physical separation of ${\sim}75\,\mathrm{kpc}$ \citep{weaver2022}, though this galaxy does not have a measured spectroscopic redshift. The closest neighbor with a spectroscopic redshift and within $1000\,\mathrm{km\,s^{-1}}$ of \name{} is at a projected physical separation of $1200\,\mathrm{kpc}$ \citep{khostovan2025}. While we cannot fully rule out a tidal origin for the observed morphology of \name{}, this shows that it does not currently have any close tidal companions. There is always the possibility of a past tidal interaction with a satellite that is no longer in the vicinity of \name{}. Stellar tidal tails should have colors that are similar to the host galaxy, particularly the outskirts of the host galaxy where stars are most easily unbound. In Fig.~\ref{fig:gal_tail_col} we plot the F115W--F444W color profile\footnote{We note that the trend in Fig.~\ref{fig:gal_tail_col} is not sensitive to the filter combination that we use to measure the galaxy/tail color.} for the main disk of \name{} compared to the tail (both for individual sources and integrated over tail region). The color profile is extracted in $0.1\arcsec$ circular annuli from the galaxy center out the the edge of the disk (${\sim}1\arcsec$). The color of the tail is clearly bluer than all regions of the main disk, thus inconsistent with the stellar emission in the tail resulting from simple tidal stripping. We cannot rule out the possibility that gas is tidally stripped from \name{} which then proceeds to form stars in situ in the tail, but gas being tidally stripped without strongly distorting the stellar distribution seems unlikely. Conversely, a stellar tail which is bluer than the main galaxy is exactly what is expected from star formation within a ram pressure stripped tail. An additional possibility is that \name{} is a collisional ring galaxy and the observed blue star-forming sources are part of a circumgalactic star-forming ring. We table further discussion of this possibility for now and leave it to be addressed comprehensively in Sect.~\ref{sec:discussion:ring}. Ultimately, given the data in hand we cannot entirely rule out this possibility, or the tidal stripping scenario for that matter, but we do favor a ram pressure stripping origin as the most-likely scenario, as we outline in the subsequent sections.

\section{Detection of an Ionized Gas Tail} \label{sec:gas_tail}

\begin{figure}[!ht]
    \centering
    \includegraphics[width=0.95\columnwidth]{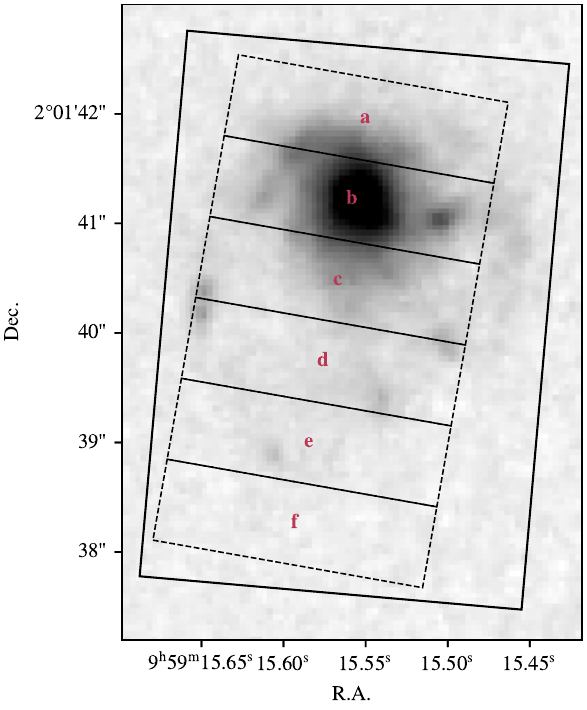}
    \caption{Overlay of the GMOS IFU-1 field-of-view (solid box) on top of the JWST F277W image of \name{}. In the dashed boxes we show the apertures along the proposed tail direction that are used for extracting spectra around the \oiifull{} doublet.}
    \label{fig:ifu_overlay}
\end{figure}

\begin{figure*}
    \centering
    \includegraphics[width=\textwidth]{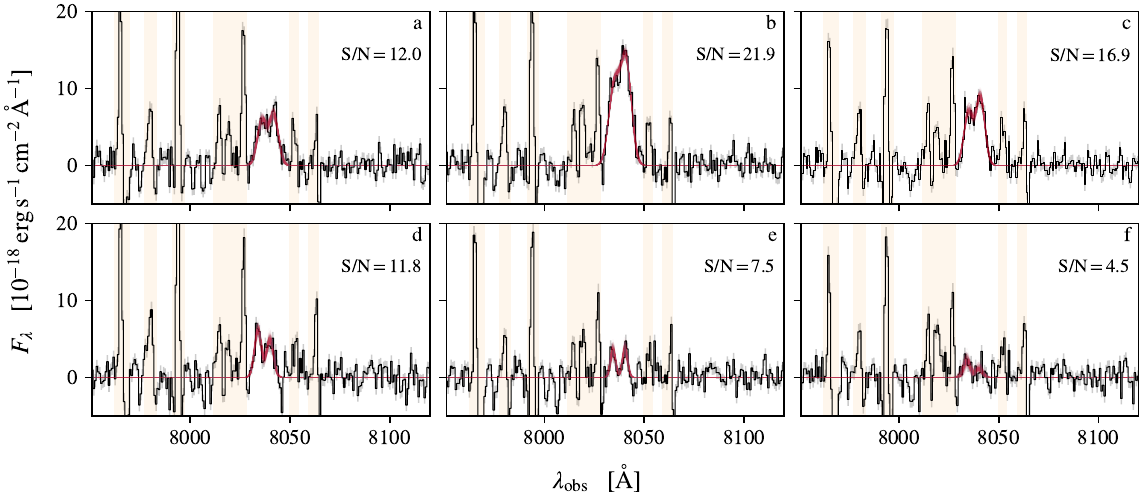}
    \caption{Extracted spectra around the \textsc{[Oii]} line for various slices along the tail axis. The labels in each panel correspond to the aperture labels in Fig.~\ref{fig:ifu_overlay}. The black line and shading show the observed spectrum and uncertainty and the red lines show 500 random samples from the posterior distribution for the best-fit double-Gaussian model. These spectra are shown post continuum subtraction with a first-order polynomial. For each spectrum we list the signal-to-noise ratio for the detection of the \oii{} doublet. The yellow bands mark regions of the spectrum where significant sky-line residuals are seen, these regions of the spectrum are masked during fitting.}
    \label{fig:oii_specs}
\end{figure*}

\begin{figure}
    \centering
    \includegraphics[width=\columnwidth]{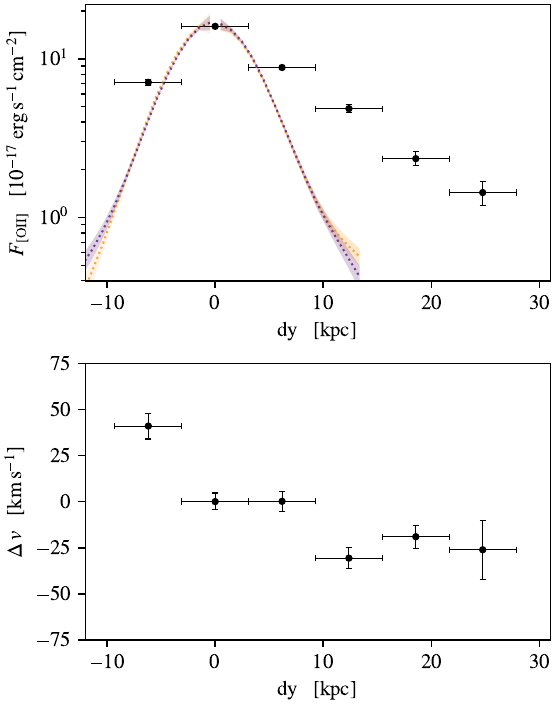}
    \caption{Profiles of \textsc{[Oii]} flux (top) and velocity (bottom) along the tail axis, the dashed vertical line marks the center of the galaxy. The $y$-errorbars correspond the the 16th and 84th percentiles from the posterior distributions for the spectral fits. The $x$-errorbars correspond to the aperture width (see Fig.~\ref{fig:ifu_overlay}). For reference, the colored lines show surface brightness profiles extracted along four quadrants for the main disk of \name{}. The yellow lines correspond to the `leading' and `trailing' sides along the axis of the tail direction and the purple lines correspond to the two perpendicular quadrants. These profiles have been scaled arbitrarily to match the normalization of the \oii{} profile. The velocity profile has been normalized such that $v=0$ corresponds to the galaxy center.}
    \label{fig:oii_prof}
\end{figure}

The association with an overdense environment is a necessary but not sufficient condition for establishing \name{} as a jellyfish galaxy undergoing RPS. Furthermore, the knots in the JWST imaging of \name{} cannot be definitively associated with the main galaxy through photometry alone. In this section we present the detection of an ionized gas tail to the south of \name{}, spatially coincident with the extra-planar rest-frame optical/near-IR sources. First we perform an aperture analysis in order to demonstrate that extra-planar ionized gas emission is present along the `tail' direction at high signal-to-noise, and second we generate maps of \textsc{[Oii]} flux and velocity in order to derive rough constraints on the spatial distribution, given our seeing-limited angular resolution.

\subsection{Aperture Analysis} \label{sec:gas_tail:aper}

In Fig.~\ref{fig:ifu_overlay} we show the GMOS IFU FOV (solid line) overlaid on the F277W image of \name{}. We also show 6 apertures along the tail axis, labeled `a' through `f', which have widths of $2.5\arcsec$ and heights of $0.75\arcsec$. The height of $0.75\arcsec$ corresponds to the FWHM image quality (${\sim}0.7\arcsec-0.8\arcsec$) and the width of $2.5\arcsec$ is chosen to span the full optical disk of \name{}. A key goal of the GMOS IFU observations of \name{} is to test for the presence of an ionized gas tail that is co-spatial with the inferred ram pressure tail from broadband imaging. To do so we extract spectra around the \oiifull{} lines for each aperture along the tail axis. If a stripped ionized gas tail is present then we should see \oii{} emission that extends beyond the main optical disk of the galaxy, and specifically this should be seen to the south of \name{}.
\par
Fig.~\ref{fig:oii_specs} show these extracted spectra. We fit each spectrum with a model that is a composite of a continuum model and a double-Gaussian model for the \oii{} doublet, given by
\begin{equation}
    F_\lambda(\lambda) = P(a_0, a_1, \lambda) + \mathcal{N}(A_1, v, \sigma, \lambda) + \mathcal{N}(A_2, v, \sigma, \lambda),
\end{equation}
\noindent
where $\lambda$ is the observed wavelength, $P$ is a first-order polynomial to account for any offset or tilt in the continuum level, and $\mathcal{N}$ are normal distributions with peak fluxes $A_1$ and $A_2$ corresponding to the two \oii{} lines. We tie both lines in the \oii{} doublet to a common velocity ($v$) and linewidth ($\sigma$). We also set a physically motivated prior range of $0.35 < A_2/A_1 < 1.5$ for the flux ratio between the two \oii{} lines \citep{copetti2002,pradhan2006}. For each extracted spectrum this gives six free parameters. We determine the best-fit parameters using the Markov-Chain Monte Carlo package \texttt{EMCEE} \citep{foreman-mackey2013} with flat priors on all model parameters. Chains are all checked for convergence by ensuring that the number of samples is greater than $50 \tau$, where $\tau$ is the chain autocorrelation time (see \href{https://emcee.readthedocs.io/en/stable/tutorials/autocorr/}{\texttt{EMCEE} documentation}).
\par
The best fit spectral models are shown in Fig.~\ref{fig:oii_specs} in red. We note that the observed spectra and best-fit model in Fig.~\ref{fig:oii_specs} have had the best-fit continuum polynomial subtracted off. The yellow bands in the panels mark regions of the spectra that are masked during fitting due to significant residuals from sky line subtraction. We calculate the total \oii{} flux as
\begin{equation}
    F_\mathrm{[OII]} = \sqrt{2 \pi}\sigma (A_1 + A_2).
\end{equation}
We detect the \oii{} doublet in all apertures. This includes detections at separations well beyond the main stellar disk and even beyond the extra-planar sources seen in the JWST imaging. In Fig.~\ref{fig:oii_prof}(top) we plot \oii{} flux as a function of distance from the galaxy center (dy) in kiloparsecs. We consider the flux profile shown in Fig.~\ref{fig:oii_prof} to be indicative of a detected ionized gas tail to the south of \name{}. The ionized gas emission extends beyond the optical galaxy disk by ${\sim}20\,\mathrm{kpc}$ all the way to the edge of the IFU field-of-view. For reference, in Fig.~\ref{fig:oii_prof}(top) we overplot F277W surface-brightness profiles along the `leading' and `trailing' sides of the main disk, as well as the two perpendicular directions. These profiles are measured after convolution to the resolution of the GMOS spectroscopy and are arbitrarily shifted to match the normalization of the \oii{} profile. These profiles reflect the symmetric nature of the galaxy disk (there is little-to-no difference between the different quadrants of the galaxy) and highlight the excess \oii{} emission that is present for $\mathrm{dy} \gtrsim 10\,\mathrm{kpc}$.
\par
In Fig.~\ref{fig:oii_prof}(bottom) we show the velocity profile for the detected ionized gas along the tail axis. A clear velocity gradient is observed from the the leading side through to the far edge of the tail. This coherence definitively associates this extra-planar ionized gas emission with the main galaxy. It also indirectly indicates that the extra-planar broadband optical/near-IR (rest frame) emission to the south of \name{} is also associated with the main galaxy.

\subsection{Ionized Gas Maps} \label{sec:gas_tail:maps}

\begin{figure*}
    \centering
    \includegraphics[width=\textwidth]{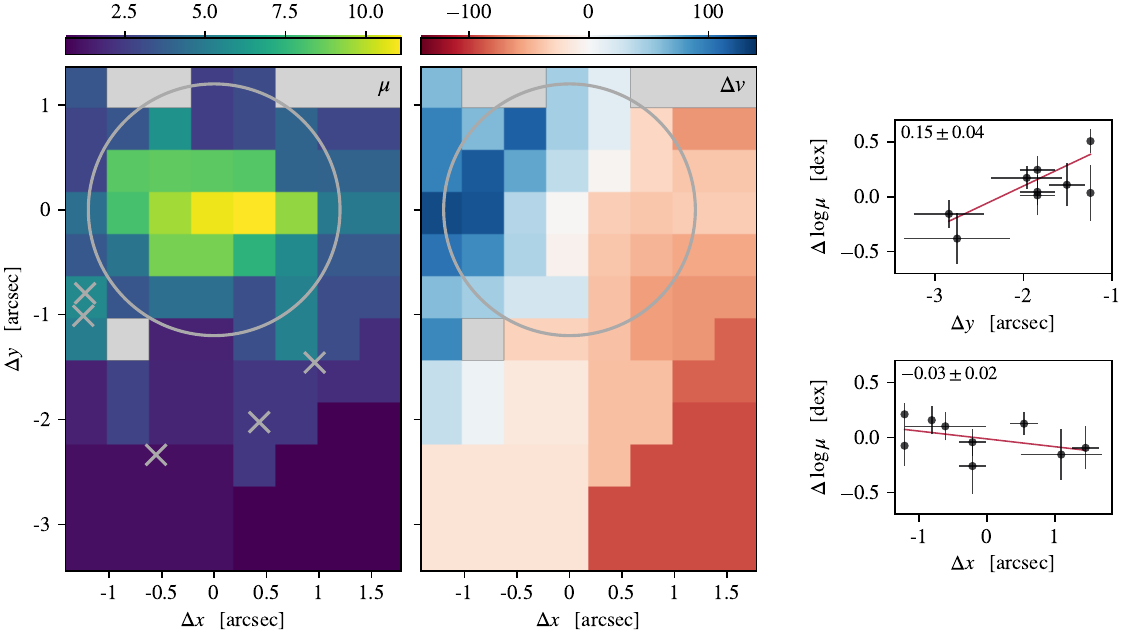}
    \caption{Maps of \oii{} surface brightness and velocity for \name{} (see Fig.~\ref{fig:ifu_overlay} for the orientation of GMOS IFU). \added{We only show spatial bins with \oii{} detected at $\mathrm{S/N} > 3$, undetected pixels/bins are shown in grey in each panel.} The grey circle in each panel shows the extent of the main stellar disk ($2\times r_e$), and the grey markers highlight the position of the JWST tail sources. Maps are in units of $10^{-17}\;\mathrm{erg\,s^{-1}\,cm^{-2}\,arcsec^{-2}}$ and $\mathrm{km\,s^{-1}}$ for surface brightness and velocity, respectively. \added{The right-hand panels show surface brightness profiles (with arbitrary normalization) along the $\Delta x$ and $\Delta y$ directions. Each panel also shows the best-fit slope and uncertainty from a linear fit.}}
    \label{fig:oii_maps}
\end{figure*}

To explore the ionized gas emission in the tail region in more detail, we now construct maps of \oii{} surface brightness and velocity. To do so we need to consider the limitations of our IFU spectroscopy both in terms of angular resolution and depth. We first bin up the IFU cube from its native pixel scale of $0.1\arcsec$ to a pixel scale of $0.4\arcsec$, in order to roughly Nyquist sample the seeing of our data. We then extract spectra from spatial bins that are determined from the PowerBin algorithm -- a modern alternative to Voronoi binning \citep{cappellari2025}. For each spaxel, we calculate the signal-to-noise ratio for the flux measured over a ${\sim}500\,\mathrm{km\,s^{-1}}$ window centered on the observed wavelength of the midpoint of the \oii{} doublet. We then bin spatially according to a target minimum $\mathrm{S/N}$ of five, though many spaxels in the galaxy disk already have signal-to-noise in excess of this target.  The binned spectra are fit with the identical model and procedure outlined in Sect.~\ref{sec:gas_tail:aper}.
\par
In Fig.~\ref{fig:oii_maps} we show the resulting \oii{} surface brightness and velocity maps. We only include spatial bins where the \oii{} flux is detected at $\mathrm{S/N} > 3$ and the velocity converges on a uni-modal posterior distribution. Furthermore, we visually inspect all of the fits to ensure that they are consistent with the \oii{} doublet being fit and that they are not obviously fitting to noise spikes or sky-line residuals. Fig.~\ref{fig:oii_maps} shows that the ionized gas emission is not limited just to the stellar sources visible in the JWST imaging, but instead fills the full region to the south of \name{}. \added{Qualitatively,} the diffuse \oii{} surface brightness falls off steadily in the $-y$-direction beyond the galaxy disk, and there is no clear structure \added{(gradient or otherwise)} across the tail in the $x$-direction. \added{These facts are supported quantitatively by surface brightness profiles along the $x$- and $y$-directions shown in Fig.~\ref{fig:oii_maps}. For the surfrace brightness profile along the $x$-direction we have removed the gradient along the $y$-direction in order to align each spatial bin to a common normalization. The slope of the best-fit profile along the $y$-direction is positive whereas the gradient along the $x$-direction is consistent with zero.}
\begin{figure}
    \centering
    \includegraphics[width=\columnwidth]{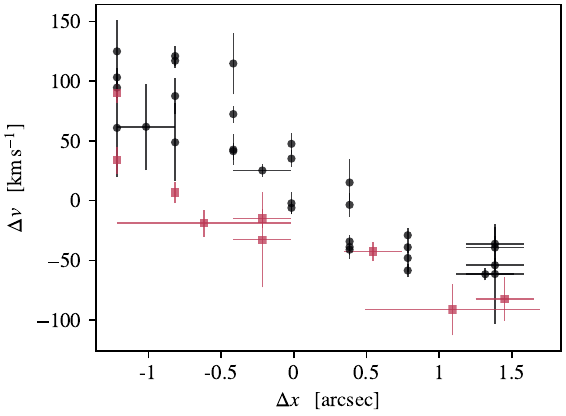}
    \caption{Position velocity diagram along the $\Delta x$-direction from Fig.~\ref{fig:oii_maps}. The main galaxy disk is shown by the black points and the tail is shown by the red points.}
    \label{fig:pvd}
\end{figure}
\par
Kinematically, the ionized gas in the disk follows a pattern of ordered rotation. There is some evidence that signatures of this rotation pattern is also present in the extraplanar gas to the south, though the spatial bins become large and the structure can only be mapped very coarsely. In Fig.~\ref{fig:pvd} we show two position-velocity diagrams (PVDs), roughly corresponding to the galaxy disk and tail. The disk pixels are defined to be those with $\Delta y \ge -1$ in Fig.~\ref{fig:oii_maps} and the tail pixels those with $\Delta y < -1$. We then extract PVDs as a function of $\Delta x$, which roughly corresponds to the plane of rotation in the disk. Both PVDs show gradients consistent with rotation, with both showing similar slopes (at least given the small number of data points for the tail). There is an offset in systemic velocity between the disk and the tail of ${\sim}50\,\mathrm{km\,s^{-1}}$, as expected from Fig.~\ref{fig:oii_prof}.
\par
This is consistent with observations of jellyfish galaxies at $z \sim 0$, where ordered ionized gas rotation is almost always observed in the disk, and this rotation pattern is often also observable in regions of the tail as the stripped gas retains some of its orbital angular momentum as it is removed from the galaxy disk \citep[e.g.][]{poggianti2017,gullieuszik2017,moretti2018_ring,lee2022_gmos_tails,luo2023}.

\section{extra-planar Star Formation} \label{sec:tail_sf}

\begin{figure*}
    \centering
    \includegraphics[width=0.495\textwidth]{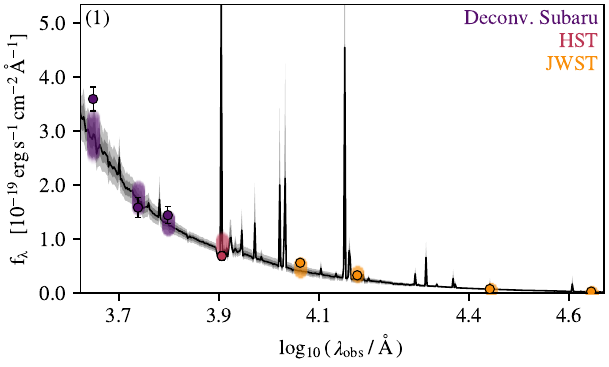}
    \includegraphics[width=0.495\textwidth]{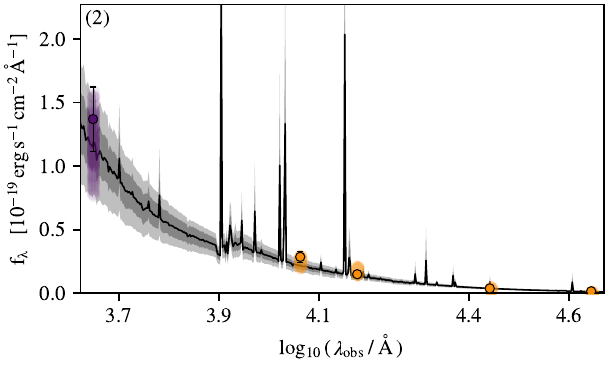}
    \includegraphics[width=0.495\textwidth]{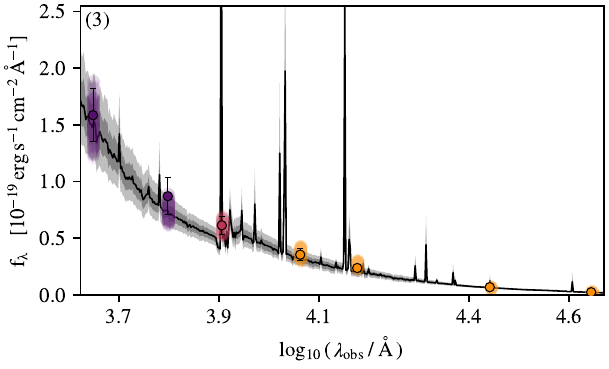}
    \includegraphics[width=0.495\textwidth]{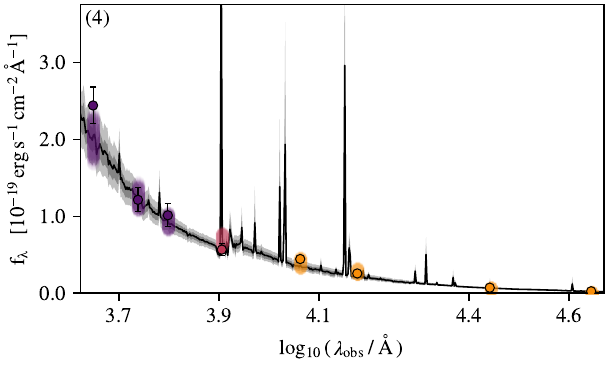}
    \caption{\texttt{BAGPIPES} SED fits for the four extra-planar sources identified in the tail of \name{}. Observed fluxes are shown by the data points. The solid black line shows the best-fit spectrum (50th percentile) and then gray shaded regions show the 68\% and 95\% credible regions. The colored bands show the 95\% credible region for the expected flux in each of our observed filters, given the fitted model spectra. The region number (see Fig.~\ref{fig:panels}) is listed in the upper-left corner.}
    \label{fig:sedfits}
\end{figure*}

\begin{figure*}
    \centering
    \includegraphics[width=0.495\textwidth]{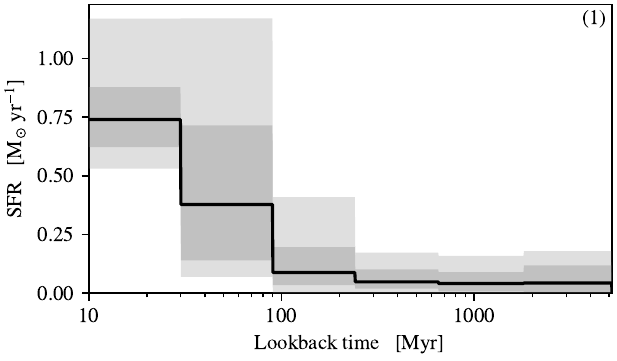}
    \includegraphics[width=0.495\textwidth]{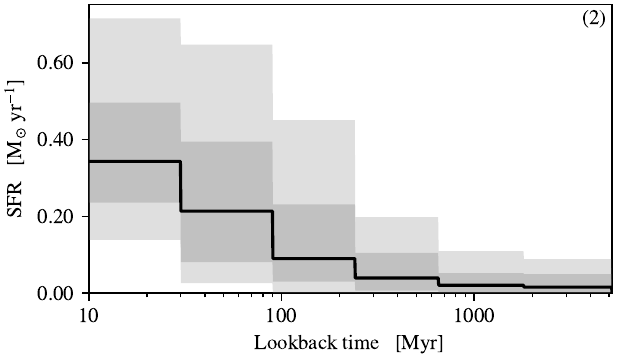}
    \includegraphics[width=0.495\textwidth]{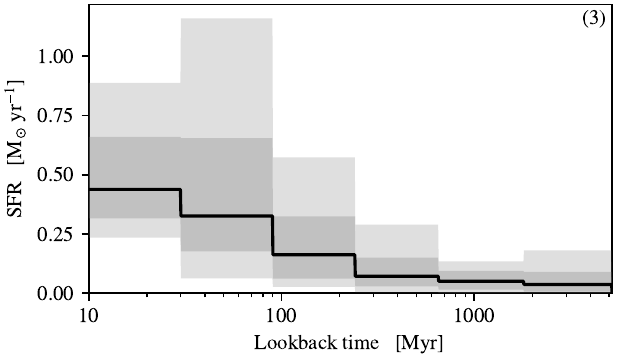}
    \includegraphics[width=0.495\textwidth]{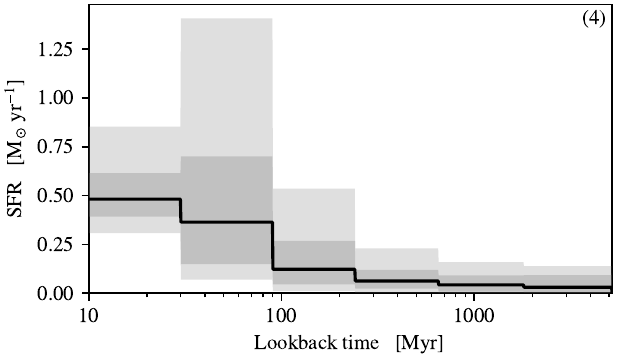}
    \caption{Star formation histories for the four extra-planar sources in the tail of \name{}. Star formation histories are determined from the SED fits shown in Fig.~\ref{fig:sedfits}. We use a flexible star formation history model with six logarithmically-spaced age bins between 0 and ${\sim}5\,\mathrm{Gyr}$. The solid black line corresponds to the median star formation history and the shaded bands show the 84th and 95th percentile credible regions.}
    \label{fig:sfh}
\end{figure*}

\begin{deluxetable}{l c c c c}
\label{tab:src_table}
\tablecaption{Properties of tail regions}
\tablehead{\colhead{ID} & \colhead{R.A.} & \colhead{Dec.} & \colhead{$\log_{10}(M_\star/\mathrm{M_\odot})$} & \colhead{SFR/$\mathrm{M_\odot\,yr^{-1}}$}}
\startdata
1 & 149.81521 & 2.02786 & $8.3_{-0.4}^{+0.2}$ & $0.46_{-0.12}^{+0.19}$ \\
2 & 149.81502 & 2.02747 & $8.0_{-0.3}^{+0.2}$ & $0.25_{-0.09}^{+0.12}$ \\
3 & 149.81476 & 2.02761 & $8.3_{-0.2}^{+0.1}$ & $0.36_{-0.11}^{+0.23}$ \\
4 & 149.81459 & 2.02775 & $8.2_{-0.2}^{+0.2}$ & $0.39_{-0.13}^{+0.21}$ \\
\enddata
\end{deluxetable}

Given the spatial coincidence between the ionized gas tail and the extra-planar JWST sources, and given the kinematic association between ionized gas tail and the main galaxy, we believe that these extra-planar optical/near-IR sources reflect sites of in situ star formation in a ram-pressure stripped tail. In this section we estimate the properties of these star-forming knots via SED fitting. We measure photometry within the four apertures shown in Fig.~\ref{fig:panels}. We note that `region 1' (and arguably `region 4') contains two peaks in the JWST imaging but we treat it as a single source since it is blended into a single source in the deconvolved Subaru imaging. The filters used are JWST F115W, F150W, F277W, and F444W, HST F814W, and Subaru B, V, $r$, and $z$. As mentioned in Sect.~\ref{sec:data:imaging}, we use the Subaru imaging from \citet{sok2022} that is deconvolved to a resolution of $0.3\arcsec$ and prior to measuring fluxes we also convolve the JWST and HST imaging to this resolution. We only include filters in the SED fitting where the flux is detected in the source aperture at ${>}5\sigma$. We determine the flux uncertainty by randomly positioning 1000 apertures (of equal size to the source aperture) across source-free regions in the image and measuring fluxes. We take the flux error to be the sigma-clipped standard deviation of the resulting aperture flux distribution. We note that the deconvolved $z$-band flux is undetected for all four sources.
\par
We perform SED fitting using the \texttt{BAGPIPES} code \citep{carnall2018,carnall2019} with the \texttt{NAUTILUS} sampler \citep{lange2023}. We assume a flexible continuity star formation history model and a Calzetti dust attenuation law \citep{calzetti2000}. We use age bins of $[0,30,90,650,1800,t_\mathrm{max}]\,\mathrm{Myr}$, where $t_\mathrm{max}$ is the age of the Universe at $z = 1.156$. These bins correspond to a minimal and maximal age bin with approximately logarithmic time-spacing in between. Dust $A_V$ is allowed to vary freely between 0 and 2 and metallicity between 0.01 and 2.5 solar. We include a nebular emission component from \texttt{CLOUDY} \citep{ferland2017} with the ionization parameter fixed to $\log_{10}(U) = -3$ and fix the redshift to the spectroscopic redshift of \name{}.
\par
In Fig.~\ref{fig:sedfits} we show the SEDs for each of the tail sources alongside the best-fit spectrum from \texttt{BAGPIPES}. Immediately by-eye it is clear that the photometry for all of the tail sources is consistent with a very young stellar population, both by the lack of a significant Balmer break and the steep UV slope. This is reinforced quantitatively in Fig.~\ref{fig:sfh} where we show the binned star formation histories from the SED fits. All sources are consistent with star formation exclusively within the past $100\,\mathrm{Myr}$. The derived star formation rates and stellar masses for each of the tail sources are listed in Table~\ref{tab:src_table}, where the listed SFRs correspond to averages over the past $100\,\mathrm{Myr}$. Stellar masses are generally around $10^8\,\mathrm{M_\odot}$ with SFRs that are a few tenths of a solar mass per year.
\par
For further reference we also derive an integrated SFR for the main disk of \name{} via SED fitting following the same procedure and using the same filters as for the tail sources (though the galaxy disk is detected in $z$-band). We measure photometry for the disk in an aperture centered on the galaxy center with a radius of $2 \times r_e$. We estimate a log stellar mass (in solar masses) of $10.24_{-0.07}^{+0.16}$ and a SFR of $68_{-18}^{+34}\,\mathrm{M_\odot\,yr^{-1}}$, which are consistent with the COSMOS2020 values of $10.30_{-0.07}^{+0.05}$ and $71_{-14}^{+22}$ \citep{weaver2022}. This means that the tail sources individually account for ${\sim}1\text{--}2\%$ of the stellar mass of the main disk and ${\sim}0.5\%$ of the SFR (averaged over the past $100\,\mathrm{Myr}$).

\section{Discussion} \label{sec:discussion}

\subsection{Could \name{} be a Collisional Ring Galaxy?} \label{sec:discussion:ring}

As the knots in the tail of \name{} are consistent with following a ring-like structure, here we will provide a relatively detailed discussion around whether the morphology of \name{} could indicate a collisional ring galaxy (CRG) as opposed to RPS. Our ultimate conclusion is that we cannot rule-out the possibility that \name{} is a CRG, but we still favor RPS as the most-likely driver of the observed stellar and ionized gas morphology for the reasons outlined below.
\par
First and foremost, we consider the structure of the ionized gas tail shown in Fig.~\ref{fig:oii_maps}. As noted in Sect.~\ref{sec:gas_tail:maps}, there is a clear decrease in \oii{} surface brightness from the galaxy disk to the edge of the IFU field along the $y$-axis, but there is no change in \oii{} surface brightness in the tail along the $x$-axis. Conversely, ionized (and atomic) gas observations of ring galaxies show strong gas emission along the ring with a strong depression inside of the ring \citep[e.g.][]{higdon1995,higdon1996,appleton1999,finkleman2011,higdon2011,deg2023,gomez-gonzalez2024}. Such a ring morphology is not at all indicated by Fig.~\ref{fig:oii_maps}. Furthermore, ionized gas emission in CRG systems tends to be very bright along the ring structure and much fainter, or undetected, in the central galaxy \citep[e.g.][]{higdon1995,appleton1999,finkleman2011,fogarty2011,gomez-gonzalez2024}. The opposite is true for \name{}. \oii{} emission peaks at the center of the galaxy and monotonically decreases along the tail direction.
\par
Kinematically, it is difficult to distinguish between a CRG and a RPS tail, given the depth and, particularly, the angular resolution of our IFU data. In both cases we expect to see ordered rotation, especially for RPS when the axis of rotation in the disk is roughly perpendicular to the tail direction -- as it is for \name{}. Some ring galaxies show kinematic misalignment, and even counter rotation, between the central galaxy and the ring \citep[e.g.][]{merrifield1994,moretti2018_ring,katkov2022,deg2023}, though this is not always the case \citep[e.g.][]{finkleman2011}. Regardless, we do not have sufficient angular resolution to determine whether or not there is kinematic misalignment between the galaxy and tail for \name{}. Though we can rule out counter rotation.
\par
In terms of stellar morphology, at the very least, \name{} does not seem to be a classical CRG with an outer ring clearly detached from the central galaxy (e.g.\ such as AM 0644-741 or Hoag's Object). Indeed, the compact knots detected by JWST can be traced back to the stellar disk via spiral arms extending from the eastern and western sides of the galaxy. It does bear mention that the Cartwheel galaxy shows spokes connecting the central galaxy to the outer ring \citep[e.g.][]{hosseinzadeh2023}. At $z \sim 0$, jellyfish galaxies show a wide range of stellar morphologies in their stripped tails. These include, linear stellar streams, largely detached compact stellar knots, as well as relatively smooth, one-sided stellar extensions \citep[e.g.][]{cortese2007,jachym2014,poggianti2025}, and \name{} fits well within this diverse set.
\par
Finally, we note that galaxies undergoing RPS and CRGs need not be mutually exclusive.  There is at least one known example of a Hoag's-like ring galaxy being subsequently ram pressure stripped, as presented in \citet{moretti2018_ring}.

\subsection{Implications for Ram Pressure Stripping} \label{sec:discussion:tail}

In this work we present evidence for the first ram-pressure stripped ionized gas tail at $z>1$, which is found trailing to the south of the starburst galaxy \name{} at $z=1.156$. We also constrain for the first time the strength of extra-planar star formation in a potential ram-pressure stripped tail at these redshifts. For this section, we make the assumption that the morphology of \name{} is being driven by RPS, and then discuss the implications. While there are no direct comparisons for \name{} at $z>1$, we can compare the observed properties of \name{} to jellyfish galaxies at lower redshift.
\par
We first address the plausibility of RPS given the overdensities that \name{} appears associated with (see Fig.~\ref{fig:lss}). The smaller (${\sim}\mathrm{few} \times 10^{13}$) halo centered directly on \name{} is, at best, marginally detected in X-rays with a luminosity of $8 \pm 6 \times 10^{42}\,\mathrm{erg\,s^{-1}}$, so while it may support a dense ICM we cannot say for sure. At this halo mass, such a system would have to be over-bright in X-rays by roughly an order of magnitude in order to be detected at the above noise level \citep[e.g.][]{sun2012,zheng2023_erosita,braspenning2025}. The large overdensity to the northeast of \name{} has an X-ray luminosity of $2 \pm 0.5 \times 10^{43}\,\mathrm{erg\,s^{-1}}$ \citep{toni2024} detected at $4\sigma$. This X-ray luminosity is not dissimilar from values for massive groups and low-mass clusters at $z \sim 0$ \citep[e.g.][]{sun2012}. Based on this X-ray luminosity we obtain an order of magnitude estimate for the corresponding ICM density of ${\sim}10^{-27}-10^{-26}\,\mathrm{g\,cm^{-3}}$ for the inner regions of the halo \citep[e.g.][]{peterson2006,sun2012}. Even if the ICM density falls off by one or two orders of magnitude towards the outskirts of the halo, this still gives possible ram pressure strengths of ${\sim}10^{-11} \text{--} 10^{-14}\,\mathrm{g\,cm^{-1}\,s^{-2}}$ for galaxy velocities between a few hundred and a thousand kilometers per second. These ram pressure strengths are similar in magnitude to values estimated from observations of RPS in the local Universe and from hydrodynamic simulations \citep[e.g][]{gunn1972,steinhauser2016,jaffe2018,roberts2019,yun2019,broderick2025}. Thus based on the X-ray data available for this system, there is no reason to exclude RPS as a perturbing force for \name{}. In fact, assuming that ICM profiles remain self-similar out to $z>1$ \citep{mcdonald2017}, then the strength of ram pressure (at fixed group/cluster mass) at these redshifts should be as strong, if not stronger, than at $z \sim 0$ (given the evolution in the critical density of the Universe). At some point this self-similarity likely breaks down at epochs where the ICM is still in the process of formation, though what redshift range this corresponds to is still uncertain.
\par
At lower redshift, \citet{boselli2019} detect striking $50\text{--}100\,\mathrm{kpc}$ ram-pressure stripped \oii{} tails behind two satellite galaxies in a ${\sim}10^{14}\,\mathrm{M_\odot}$ cluster at $z \sim 0.7$. \citeauthor{boselli2019} find a typical \oii{} surface brightness in the stripped tails of a few times $10^{-18}\,\mathrm{erg\,s^{-1}\,cm^{-2}\,arcsec^{-2}}$, whereas we find a typical \oii{} surface brightness of ${\sim}10^{-17}\,\mathrm{erg\,s^{-1}\,cm^{-2}\,arcsec^{-2}}$. The observed size of tail that we detect is far shorter than the tails found by \citet{boselli2019}, though for galaxy 345 in \citeauthor{boselli2019} the \oii{} surface brightness within $10\text{--}20\,\mathrm{kpc}$ of the galaxy disk is close to the typical value that we find for \name{}. This may suggest that the intrinsic length of the ionized gas tail behind \name{} is much longer than we observe, but this will require significantly deeper observations to constrain. At lower redshift, the largest compilation of \oii{} detected ram-pressure stripped tails is \citet{moretti2022} with ${\sim10}$ RPS galaxies in the A2744 and A370 clusters at $z \sim 0.3\text{--}0.4$. A new revelation from \citeauthor{moretti2022} is the fact that the ram-pressure tails in their sample show high \oii{}-to-$\mathrm{H\alpha}$ ratios of 2--3, compared to typical galaxy disks with \oii{}-to-$\mathrm{H\alpha}$ ratios less than unity. They argue that this anomalous ratio is either driven by the low gas densities in ram pressure tails or is due to the complex interaction between the stripped tail and the surrounding ICM.  At $z=1.156$, the $\mathrm{H\alpha}$ line falls between the J- and H-bands and is thus unobservable from the ground for \name{}. So while we do not have constraints on the $\mathrm{H\alpha}$ emission for \name{}, these high \oii{}-to-$\mathrm{H\alpha}$ ratios seen at intermediate redshifts may be aiding our ability to detect \oii{} in the tail of \name{} at $z>1$. Moving forward, constraints on the $\mathrm{H\alpha}$ emission from \name{} and its stripped tail will require spectroscopy from space.
\par
In Fig.~\ref{fig:panels} we highlight four extra-planar regions in the tail of \name{} that are clear from the JWST imaging. We note that regions 1 and 4 are resolved into two sources in the JWST imaging but are blended into one in the deconvolved Subaru imaging that we use for SED fitting. These regions are blue in color and are consistent with exclusively young stellar populations formed within the past $100\,\mathrm{Myr}$. Detached star-forming knots in Virgo cluster ram-pressure tails tend to have stellar masses of ${\sim}10^3\text{--}10^5\,\mathrm{M_\odot}$ and SFRs of ${\sim}10^{-4}\text{--}10^{-3}\,\mathrm{M_\odot\,yr^{-1}}$ \citep{hester2010,fumagalli2011,boselli2021_ic3476,junais2021,boselli2022_review}. For star-forming knots in the tails of jellyfish galaxies from the GASP survey (GAs Stripping Phenomena in galaxies with MUSE, \citealt{poggianti2017}), stellar masses range between ${\sim}10^6\text{--}10^8\,\mathrm{M_\odot}$ and SFRs between ${\sim}10^{-4}\text{--}10^{-1}\,\mathrm{M_\odot\,yr^{-1}}$. These can be compared to the typical stellar masses of ${\sim}10^8\,\mathrm{M_\odot}$ and SFRs of ${\sim}0.3\,\mathrm{M_\odot\,yr^{-1}}$ derived in this work for the four tail regions. These differences relative to the knots for Virgo galaxies is likely driven by the differences in physical resolution such that the JWST sources in regions in this work may correspond to multiple small sources blended together. Relative to $z \sim 0$, at $z \sim 1$ there is an increase by an order of magnitude in the normalization of the SFMS \citep[e.g.][]{speagle2014} which may also manifest itself in the specific star formation rate of detached star-forming knots. Relative to GASP sources, the physical properties of the four tail regions in this work are roughly consistent with the most-massive GASP knots, though the SFRs for the regions in the tail of \name{} are slightly larger \citep{poggianti2019_extraplanar_sf}. Incorporating the factor of ${\sim}10$ evolution in the SFMS moves the sources from this work well within the scatter of of the GASP sample, though it remains to be seen whether this factor applies to star formation in stripped tails.
\par
We can speculate on the fate of these extra-planar star-forming knots observed at $z > 1$. They may remain bound to the dark-matter halo of \name{} and eventually be re-accreted on to the disk. The radial velocity offset of the tail relative to the main galaxy is modest at $\lesssim 50\,\mathrm{km\,s^{-1}}$, but we cannot say whether this is indicative of a modest 3D velocity offset or a larger 3D velocity offset that is primarily in the plane of the sky. Simulations of isolated RPS find that most stars formed within $20\,\mathrm{kpc}$ of the disk will eventually fall back onto the galaxy \citep{tonnesen2012,akerman2025}, which is about the maximum separation that we observe for tail sources from the disk of \name{}. Thus, at least based on simulations, this may be the most likely fate. Alternatively these knots are unbound from \name{}. In this scenario there are two broad possibilities, either they survive as a compact stellar collection or they are tidally disrupted and disperse. The latter would contribute to the building up the intracluster light (ICL) that is observed in clusters from $z \sim 0$ out to $z \gtrsim 1$ \citep[e.g.][]{lin2004,jimenez-teja2019,burke2012,joo2023}. Sect.~\ref{sec:tail_sf} suggests that ${\sim}10^8\text{--}10^9\,\mathrm{M_\odot}$ of total stellar mass has formed in the wake of \name{} over the past ${\sim}100\,\mathrm{Myr}$. Exactly how much ICL could be built up in this fashion over the history of the cluster will depend enormously on the statistical frequency of RPS, and specifically extra-planar star formation in ram-pressure tails. Hydrodynamic simulations predict that up to 30\% of the ICL is built-up by `intracluster star formation' in gas clouds stripped from infalling galaxies \citep{puchwein2010}, which is analogous to the extra-planar star formation that we measure for \name{}. However, some simulation results suggest that this extra-planar star formation plays a much smaller role in ICL accumulation \citep{tonnesen2012,akerman2025}. The other possibility is that these sources are self-gravitating and remain as bound objects that become separated from \name{}. Such objects would, by definition, be dark-matter deficient. This notion of small `galaxies' forming within ram pressure tails is not a new one and has been speculated on both via the lens of low-redshift observations as well as hydrodynamic simulations \citep[e.g.][]{poggianti2019_extraplanar_sf,lora2024}. The masses of the tail sources in this work are consistent with the masses of dwarf galaxies in clusters at $z \sim 0$, but we have limited information on their sizes. The SED fits in Sect.~\ref{sec:tail_sf} are done at $0.3\arcsec$ resolution (limited by the deconvolved Subaru images), but it is clear from JWST images in Fig.~\ref{fig:panels} that the continuum sizes of these sources are smaller than this.  We estimate intrinsic FWHM sizes for these tail sources of ${\sim}0.13\arcsec \approx 1\,\mathrm{kpc}$ (see Appendix~\ref{app:src_size}), corresponding to effective radii of ${\sim}500\,\mathrm{pc}$. Thus the size and stellar mass of continuum sources in the tail of \name{} do coincide with typical values for dwarf galaxies in low-$z$ clusters \citep[e.g.][]{kormendy2009, marleau2025}, but the dynamical evolution of these sources from $z \sim 1$ to the present-day is highly uncertain. Even if they do survive as bound objects, which is far from a given, they may end up looking very different than they do at $z>1$.
\par
Finally, we briefly discuss the current evidence for RPS at $z>1$, beyond this work. \citet{xu2025} present compelling evidence for ram-pressure stripped molecular gas tails in a few galaxies\footnote{Here our discussion is primarily referring to galaxies `D4', `D5', and `D6' from \citet{xu2025} which seem to be the most compelling examples.} that are part of a $z=2.51$ protocluster with a mass of ${\sim}10^{14}\,\mathrm{M_\odot}$. If confirmed, this would show that RPS is truly operating all the way back to Cosmic Noon. The galaxies from \citet{xu2025} are broadly of a similar stellar mass and SFR to \name{}, and like this work, the proposed ram-pressure tails are shown to be kinematically connected to the main galaxy. We do not have molecular gas observations of \name{}, and \citet{xu2025} do not have resolved rest-frame optical spectroscopy, so the direct comparisons are limited. That said, these works highlight the potential multiwavelength nature of ram-pressure tails at $z>1$, a fact that is already well known in the low-redshift Universe \citep[e.g.][]{chung2007,ebeling2014,poggianti2017,boselli2018,jachym2019,roberts2021_LOFARclust,edler2023,souchereau2025}. Moving forward it will be extremely valuable to obtain multi-phase observations of these high-redshift ram pressure tails to further understand this process in the early Universe and make detailed comparisons to examples at late times. There is also more marginal evidence for RPS at $z>1$ in the literature, which has often been limited by angular resolution. \citet{noble2019} present possible molecular gas extensions for cluster galaxies at $z \sim 1.6$, which may be tracing RPS. More indirect inferences have also been made in favor RPS in this redshift range  \citep[e.g.][]{nantais2017,hamadouche2024}, though in these cases it is more difficult conclusively associate the observed trends with specific quenching mechanisms.

\section{Conclusions} \label{sec:conclusions}

In this work we present \name{} as a candidate jellyfish galaxy undergoing ram pressure stripping at $z=1.156$. Given the paucity of direct evidence for ram pressure stripping at $z > 1$, \name{} is represents an important new addition to the broader understanding of environmental quenching at these early times. The key takeaways from this work are the following:
\begin{itemize}
    \item[-] \name{} displays a relatively symmetric stellar disk accompanied by a trail of star-forming knots to the south of the main galaxy, as identified from multi-band JWST imaging.
    
    \item[-] We detect an ionized gas tail detected via the \oiifull{} doublet that is co-spatial with these extra-planar continuum sources and kinematically coherent with the main galaxy disk.
    
    \item[-] The star-forming knots in the tail have stellar masses of ${\sim}10^8\,\mathrm{M_\odot}$ and SFRs of ${\sim}0.3\,\mathrm{M_\odot\,yr^{-1}}$, though it is possible that the values represent to totals for smaller knots below our resolution level that are being blended together. In total, these sources account for roughly $1\%$ of the stellar mass and SFR of the main galaxy disk.
\end{itemize}
\noindent
This work highlights the powerful combination of depth and angular resolution afforded by the \textit{James Webb Space Telescope}. This permits spatially resolved analyses in order to directly constrain the quenching mechanisms at work in the high-redshift Universe. \name{} is now an important laboratory in this regard and efforts moving forward will be dedicated to confirming the nature of this galaxy via multi-wavelength observations of the candidate ram pressure tail presented in this work.

%% Also note that the akcnowlodgment environment does not support long amounts of text. If you have a lot of people and institutions to acknowledge, do not use this command. Instead, create a new \section{Acknowledgments}.
\begin{acknowledgments}
IDR acknowledges support from the Banting Fellowship Program. We thanks the COSMOS-Web team for providing publicly available, science-ready data products. This work was enabled by observations made from the Gemini North telescope, located within the Maunakea Science Reserve and adjacent to the summit of Maunakea. We are grateful for the privilege of observing the Universe from a place that is unique in both its astronomical quality and its cultural significance. This work is based in part on observations obtained at the international Gemini Observatory, a program of NSF NOIRLab, which is managed by the Association of Universities for Research in Astronomy (AURA) under a cooperative agreement with the U.S. National Science Foundation on behalf of the Gemini Observatory partnership: the U.S. National Science Foundation (United States), National Research Council (Canada), Agencia Nacional de Investigaci\'{o}n y Desarrollo (Chile), Ministerio de Ciencia, Tecnolog\'{i}a e Innovaci\'{o}n (Argentina), Minist\'{e}rio da Ci\^{e}ncia, Tecnologia, Inova\c{c}\~{o}es e Comunica\c{c}\~{o}es (Brazil), and Korea Astronomy and Space Science Institute (Republic of Korea). This work is based in part on observations made with the NASA/ESA/CSA James Webb Space Telescope. The data were obtained from the Mikulski Archive for Space Telescopes at the Space Telescope Science Institute, which is operated by the Association of Universities for Research in Astronomy, Inc., under NASA contract NAS 5-03127 for JWST. This research is based in part on observations made with the NASA/ESA Hubble Space Telescope obtained from the Space Telescope Science Institute, which is operated by the Association of Universities for Research in Astronomy, Inc., under NASA contract NAS 5–26555.
\end{acknowledgments}


\vspace{5mm}
\facilities{Gemini:North (GMOS), HST (ACS), JWST (NIRCam), Subaru (Suprime-Cam)}


\software{AstroPy \citep{astropy2013}, Bagpipes \citep{carnall2018}, Matplotlib \citep{hunter2007}, Nautilus \citep{lange2023}, NumPy \citep{harris2020}}

\bibliography{main}{}

@string{aj = {AJ}}

@string{apj = {ApJ}}

@string{apjl = {ApJL}}

@string{apjs = {ApJS}}

@string{araa = {ARA{\&}A}}

@string{mnras = {MNRAS}}

@string{nat = {Nature}}

@string{pasp = {PASP}}

@string{pasa = {PASA}}

@ARTICLE{akerman2025,
       author = {{Akerman}, Nina and {Tonnesen}, Stephanie and {Poggianti}, Bianca M. and {Smith}, Rory and {Werle}, Ariel and {Giunchi}, Eric and {Vulcani}, Benedetta and {Fritz}, Jacopo},
        title = "{What goes around comes around: the fate of stars in stripped tails of gas}",
      journal = {arXiv e-prints},
         year = 2025,
        month = apr,
          eid = {arXiv:2504.11526},
        pages = {arXiv:2504.11526},
          doi = {10.48550/arXiv.2504.11526},
archivePrefix = {arXiv},
       eprint = {2504.11526},
 primaryClass = {astro-ph.GA},
       adsurl = {https://ui.adsabs.harvard.edu/abs/2025arXiv250411526A}
}

@article{appleton1999,
   title={Plasma and Warm Dust in the Collisional Ring Galaxy VII Zw 466 from VLA andISOObservations},
   volume={527},
   ISSN={1538-4357},
   url={http://dx.doi.org/10.1086/308074},
   DOI={10.1086/308074},
   number={1},
   journal={The Astrophysical Journal},
   publisher={American Astronomical Society},
   author={Appleton, P. N. and Charmandaris, V. and Horellou, C. and Mirabel, I. F. and Ghigo, F. and Higdon, J. L. and Lord, S.},
   year={1999},
   month=dec, pages={143–153} }

@ARTICLE{astropy2013,
   author = {{Astropy Collaboration} and {Robitaille}, T.~P. and {Tollerud}, E.~J. and
	{Greenfield}, P. and {Droettboom}, M. and {Bray}, E. and {Aldcroft}, T. and
	{Davis}, M. and {Ginsburg}, A. and {Price-Whelan}, A.~M. and
	{Kerzendorf}, W.~E. and {Conley}, A. and {Crighton}, N. and
	{Barbary}, K. and {Muna}, D. and {Ferguson}, H. and {Grollier}, F. and
	{Parikh}, M.~M. and {Nair}, P.~H. and {Unther}, H.~M. and {Deil}, C. and
	{Woillez}, J. and {Conseil}, S. and {Kramer}, R. and {Turner}, J.~E.~H. and
	{Singer}, L. and {Fox}, R. and {Weaver}, B.~A. and {Zabalza}, V. and
	{Edwards}, Z.~I. and {Azalee Bostroem}, K. and {Burke}, D.~J. and
	{Casey}, A.~R. and {Crawford}, S.~M. and {Dencheva}, N. and
	{Ely}, J. and {Jenness}, T. and {Labrie}, K. and {Lim}, P.~L. and
	{Pierfederici}, F. and {Pontzen}, A. and {Ptak}, A. and {Refsdal}, B. and
	{Servillat}, M. and {Streicher}, O.},
    title = "{Astropy: A community Python package for astronomy}",
  journal = {A\&A},
archivePrefix = "arXiv",
   eprint = {1307.6212},
 primaryClass = "astro-ph.IM",
     year = 2013,
    month = oct,
   volume = 558,
      eid = {A33},
    pages = {A33},
      doi = {10.1051/0004-6361/201322068},
   adsurl = {http://adsabs.harvard.edu/abs/2013A%26A...558A..33A}
}

@ARTICLE{balogh1999,
   author = {{Balogh}, M.~L. and {Morris}, S.~L. and {Yee}, H.~K.~C. and
	{Carlberg}, R.~G. and {Ellingson}, E.},
    title = "{Differential Galaxy Evolution in Cluster and Field Galaxies at z\~{}0.3}",
  journal = {ApJ},
   eprint = {astro-ph/9906470},
     year = 1999,
    month = dec,
   volume = 527,
    pages = {54-79},
      doi = {10.1086/308056},
   adsurl = {http://adsabs.harvard.edu/abs/1999ApJ...527...54B}
}

@ARTICLE{bellhouse2021,
       author = {{Bellhouse}, Callum and {McGee}, Sean L. and {Smith}, Rory and {Poggianti}, Bianca M. and {Jaff{\'e}}, Yara L. and {Kraljic}, Katarina and {Franchetto}, Andrea and {Fritz}, Jacopo and {Vulcani}, Benedetta and {Tonnesen}, Stephanie and {Roediger}, Elke and {Moretti}, Alessia and {Gullieuszik}, Marco and {Shin}, Jihye},
        title = "{GASP XXIX - unwinding the arms of spiral galaxies via ram-pressure stripping}",
      journal = {\mnras},
         year = 2021,
        month = jan,
       volume = {500},
       number = {1},
        pages = {1285-1312},
          doi = {10.1093/mnras/staa3298},
archivePrefix = {arXiv},
       eprint = {2010.09733},
 primaryClass = {astro-ph.GA},
       adsurl = {https://ui.adsabs.harvard.edu/abs/2021MNRAS.500.1285B}
}

@ARTICLE{blanton2009,
   author = {{Blanton}, M.~R. and {Moustakas}, J.},
    title = "{Physical Properties and Environments of Nearby Galaxies}",
  journal = {ARAA},
archivePrefix = "arXiv",
   eprint = {0908.3017},
     year = 2009,
    month = sep,
   volume = 47,
    pages = {159-210},
      doi = {10.1146/annurev-astro-082708-101734},
   adsurl = {http://adsabs.harvard.edu/abs/2009ARA%26A..47..159B}
}

@ARTICLE{boselli2006,
   author = {{Boselli}, A. and {Gavazzi}, G.},
    title = "{Environmental Effects on Late-Type Galaxies in Nearby Clusters}",
  journal = {PASP},
   eprint = {astro-ph/0601108},
     year = 2006,
    month = apr,
   volume = 118,
    pages = {517-559},
      doi = {10.1086/500691},
   adsurl = {http://adsabs.harvard.edu/abs/2006PASP..118..517B}
}

@ARTICLE{boselli2018,
       author = {{Boselli}, A. and {Fossati}, M. and {Ferrarese}, L. and {Boissier}, S. and
         {Consolandi}, G. and {Longobardi}, A. and {Amram}, P. and {Balogh}, M. and
         {Barmby}, P. and {Boquien}, M. and {Boulanger}, F. and {Braine}, J. and
         {Buat}, V. and {Burgarella}, D. and {Combes}, F. and {Contini}, T. and
         {Cortese}, L. and {C{\^o}t{\'e}}, P. and {C{\^o}t{\'e}}, S. and {Cuilland
        re}, J.~C. and {Drissen}, L. and {Epinat}, B. and {Fumagalli}, M. and
         {Gallagher}, S. and {Gavazzi}, G. and {Gomez-Lopez}, J. and {Gwyn}, S. and
         {Harris}, W. and {Hensler}, G. and {Koribalski}, B. and {Marcelin}, M. and
         {McConnachie}, A. and {Miville-Deschenes}, M.~A. and {Navarro}, J. and
         {Patton}, D. and {Peng}, E.~W. and {Plana}, H. and {Prantzos}, N. and
         {Robert}, C. and {Roediger}, J. and {Roehlly}, Y. and {Russeil}, D. and
         {Salome}, P. and {Sanchez-Janssen}, R. and {Serra}, P. and
         {Spekkens}, K. and {Sun}, M. and {Taylor}, J. and {Tonnesen}, S. and
         {Vollmer}, B. and {Willis}, J. and {Wozniak}, H. and {Burdullis}, T. and
         {Devost}, D. and {Mahoney}, B. and {Manset}, N. and {Petric}, A. and
         {Prunet}, S. and {Withington}, K.},
        title = "{A Virgo Environmental Survey Tracing Ionised Gas Emission (VESTIGE). I. Introduction to the survey}",
      journal = {A\&A},
         year = "2018",
        month = "Jun",
       volume = {614},
          eid = {A56},
        pages = {A56},
          doi = {10.1051/0004-6361/201732407},
archivePrefix = {arXiv},
       eprint = {1802.02829},
 primaryClass = {astro-ph.GA},
       adsurl = {https://ui.adsabs.harvard.edu/abs/2018A&A...614A..56B}
}

@ARTICLE{boselli2022_review,
       author = {{Boselli}, Alessandro and {Fossati}, Matteo and {Sun}, Ming},
        title = "{Ram pressure stripping in high-density environments}",
      journal = {A\&AR},
         year = 2022,
        month = dec,
       volume = {30},
       number = {1},
          eid = {3},
        pages = {3},
          doi = {10.1007/s00159-022-00140-3},
archivePrefix = {arXiv},
       eprint = {2109.13614},
 primaryClass = {astro-ph.GA},
       adsurl = {https://ui.adsabs.harvard.edu/abs/2022A&ARv..30....3B}
}

@ARTICLE{boselli2021_ic3476,
       author = {{Boselli}, A. and {Lupi}, A. and {Epinat}, B. and {Amram}, P. and {Fossati}, M. and {Anderson}, J.~P. and {Boissier}, S. and {Boquien}, M. and {Consolandi}, G. and {C{\^o}t{\'e}}, P. and {Cuillandre}, J.~C. and {Ferrarese}, L. and {Galbany}, L. and {Gavazzi}, G. and {G{\'o}mez-L{\'o}pez}, J.~A. and {Gwyn}, S. and {Hensler}, G. and {Hutchings}, J. and {Kuncarayakti}, H. and {Longobardi}, A. and {Peng}, E.~W. and {Plana}, H. and {Postma}, J. and {Roediger}, J. and {Roehlly}, Y. and {Schimd}, C. and {Trinchieri}, G. and {Vollmer}, B.},
        title = "{A Virgo Environmental Survey Tracing Ionised Gas Emission (VESTIGE). IX. The effects of ram pressure stripping down to the scale of individual HII regions in the dwarf galaxy IC 3476}",
      journal = {A\&A},
         year = 2021,
        month = feb,
       volume = {646},
          eid = {A139},
        pages = {A139},
          doi = {10.1051/0004-6361/202039046},
archivePrefix = {arXiv},
       eprint = {2012.07377},
 primaryClass = {astro-ph.GA},
       adsurl = {https://ui.adsabs.harvard.edu/abs/2021A&A...646A.139B}
}

@ARTICLE{boselli2019,
       author = {{Boselli}, A. and {Epinat}, B. and {Contini}, T. and {Abril-Melgarejo}, V. and {Boogaard}, L.~A. and {Pointecouteau}, E. and {Ventou}, E. and {Brinchmann}, J. and {Carton}, D. and {Finley}, H. and {Michel-Dansac}, L. and {Soucail}, G. and {Weilbacher}, P.~M.},
        title = "{Evidence for ram-pressure stripping in a cluster of galaxies at z = 0.7}",
      journal = {\aap},
         year = 2019,
        month = nov,
       volume = {631},
          eid = {A114},
        pages = {A114},
          doi = {10.1051/0004-6361/201936133},
archivePrefix = {arXiv},
       eprint = {1909.05491},
 primaryClass = {astro-ph.GA},
       adsurl = {https://ui.adsabs.harvard.edu/abs/2019A&A...631A.114B}
}

@ARTICLE{braspenning2025,
       author = {{Braspenning}, Joey and {Schaye}, Joop and {Pillepich}, Annalisa and {Nelson}, Dylan},
        title = "{The origin of scatter in the X-ray luminosity - halo mass relation of galaxy clusters}",
      journal = {arXiv e-prints},
         year = 2025,
        month = nov,
          eid = {arXiv:2511.09649},
        pages = {arXiv:2511.09649},
          doi = {10.48550/arXiv.2511.09649},
archivePrefix = {arXiv},
       eprint = {2511.09649},
 primaryClass = {astro-ph.CO},
       adsurl = {https://ui.adsabs.harvard.edu/abs/2025arXiv251109649B}
}

@ARTICLE{broderick2025,
       author = {{Broderick}, Ariel O. and {Roberts}, Ian D. and {Hudson}, Michael J.},
        title = "{Truncated star formation and ram pressure stripping in the Coma Cluster}",
      journal = {arXiv e-prints},
         year = 2025,
        month = may,
          eid = {arXiv:2505.10633},
        pages = {arXiv:2505.10633},
          doi = {10.48550/arXiv.2505.10633},
archivePrefix = {arXiv},
       eprint = {2505.10633},
 primaryClass = {astro-ph.GA},
       adsurl = {https://ui.adsabs.harvard.edu/abs/2025arXiv250510633B}
}

@ARTICLE{brown2023,
       author = {{Brown}, Toby and {Roberts}, Ian D. and {Thorp}, Mallory and {Ellison}, Sara L. and {Zabel}, Nikki and {Wilson}, Christine D. and {Bah{\'e}}, Yannick M. and {Bisaria}, Dhruv and {Bolatto}, Alberto D. and {Boselli}, Alessandro and {Chung}, Aeree and {Cortese}, Luca and {Catinella}, Barbara and {Davis}, Timothy A. and {Jim{\'e}nez-Donaire}, Mar{\'\i}a J. and {Lagos}, Claudia D.~P. and {Lee}, Bumhyun and {Parker}, Laura C. and {Smith}, Rory and {Spekkens}, Kristine and {Stevens}, Adam R.~H. and {Villanueva}, Vicente and {Watts}, Adam B.},
        title = "{VERTICO. VII. Environmental Quenching Caused by the Suppression of Molecular Gas Content and Star Formation Efficiency in Virgo Cluster Galaxies}",
      journal = {\apj},
         year = 2023,
        month = oct,
       volume = {956},
       number = {1},
          eid = {37},
        pages = {37},
          doi = {10.3847/1538-4357/acf195},
archivePrefix = {arXiv},
       eprint = {2308.10943},
 primaryClass = {astro-ph.GA},
       adsurl = {https://ui.adsabs.harvard.edu/abs/2023ApJ...956...37B}
}

@ARTICLE{bureau2002,
       author = {{Bureau}, M. and {Carignan}, C.},
        title = "{Environment, Ram Pressure, and Shell Formation in Holmberg II}",
      journal = {AJ},
         year = 2002,
        month = mar,
       volume = {123},
       number = {3},
        pages = {1316-1333},
          doi = {10.1086/338899},
archivePrefix = {arXiv},
       eprint = {astro-ph/0112325},
 primaryClass = {astro-ph},
       adsurl = {https://ui.adsabs.harvard.edu/abs/2002AJ....123.1316B}
}

@ARTICLE{burke2012,
       author = {{Burke}, Claire and {Collins}, Chris A. and {Stott}, John P. and {Hilton}, Matt},
        title = "{Measurement of the intracluster light at z {\ensuremath{\sim}} 1}",
      journal = {\mnras},
         year = 2012,
        month = sep,
       volume = {425},
       number = {3},
        pages = {2058-2068},
          doi = {10.1111/j.1365-2966.2012.21555.x},
archivePrefix = {arXiv},
       eprint = {1206.4735},
 primaryClass = {astro-ph.CO},
       adsurl = {https://ui.adsabs.harvard.edu/abs/2012MNRAS.425.2058B}
}

@ARTICLE{calzetti2000,
       author = {{Calzetti}, Daniela and {Armus}, Lee and {Bohlin}, Ralph C. and {Kinney}, Anne L. and {Koornneef}, Jan and {Storchi-Bergmann}, Thaisa},
        title = "{The Dust Content and Opacity of Actively Star-forming Galaxies}",
      journal = {\apj},
         year = 2000,
        month = apr,
       volume = {533},
       number = {2},
        pages = {682-695},
          doi = {10.1086/308692},
archivePrefix = {arXiv},
       eprint = {astro-ph/9911459},
 primaryClass = {astro-ph},
       adsurl = {https://ui.adsabs.harvard.edu/abs/2000ApJ...533..682C}
}

@ARTICLE{cappellari2025,
       author = {{Cappellari}, Michele},
        title = "{PowerBin: fast adaptive data binning with Centroidal Power Diagrams}",
      journal = {\mnras},
         year = 2025,
        month = dec,
       volume = {544},
       number = {2},
        pages = {1432-1446},
          doi = {10.1093/mnras/staf1726},
archivePrefix = {arXiv},
       eprint = {2509.06903},
 primaryClass = {astro-ph.IM},
       adsurl = {https://ui.adsabs.harvard.edu/abs/2025MNRAS.544.1432C}
}

@ARTICLE{carnall2018,
       author = {{Carnall}, A.~C. and {McLure}, R.~J. and {Dunlop}, J.~S. and
         {Dav{\'e}}, R.},
        title = "{Inferring the star formation histories of massive quiescent galaxies with BAGPIPES: evidence for multiple quenching mechanisms}",
      journal = {MNRAS},
         year = 2018,
        month = nov,
       volume = {480},
       number = {4},
        pages = {4379-4401},
          doi = {10.1093/mnras/sty2169},
archivePrefix = {arXiv},
       eprint = {1712.04452},
 primaryClass = {astro-ph.GA},
       adsurl = {https://ui.adsabs.harvard.edu/abs/2018MNRAS.480.4379C}
}

@ARTICLE{carnall2019,
       author = {{Carnall}, Adam C. and {Leja}, Joel and {Johnson}, Benjamin D. and
         {McLure}, Ross J. and {Dunlop}, James S. and {Conroy}, Charlie},
        title = "{How to Measure Galaxy Star Formation Histories. I. Parametric Models}",
      journal = {ApJ},
         year = 2019,
        month = mar,
       volume = {873},
       number = {1},
          eid = {44},
        pages = {44},
          doi = {10.3847/1538-4357/ab04a2},
archivePrefix = {arXiv},
       eprint = {1811.03635},
 primaryClass = {astro-ph.GA},
       adsurl = {https://ui.adsabs.harvard.edu/abs/2019ApJ...873...44C}
}

@ARTICLE{casey2023,
       author = {{Casey}, Caitlin M. and {Kartaltepe}, Jeyhan S. and {Drakos}, Nicole E. and {Franco}, Maximilien and {Harish}, Santosh and {Paquereau}, Louise and {Ilbert}, Olivier and {Rose}, Caitlin and {Cox}, Isabella G. and {Nightingale}, James W. and {Robertson}, Brant E. and {Silverman}, John D. and {Koekemoer}, Anton M. and {Massey}, Richard and {McCracken}, Henry Joy and {Rhodes}, Jason and {Akins}, Hollis B. and {Allen}, Natalie and {Amvrosiadis}, Aristeidis and {Arango-Toro}, Rafael C. and {Bagley}, Micaela B. and {Bongiorno}, Angela and {Capak}, Peter L. and {Champagne}, Jaclyn B. and {Chartab}, Nima and {Ch{\'a}vez Ortiz}, {\'O}scar A. and {Chworowsky}, Katherine and {Cooke}, Kevin C. and {Cooper}, Olivia R. and {Darvish}, Behnam and {Ding}, Xuheng and {Faisst}, Andreas L. and {Finkelstein}, Steven L. and {Fujimoto}, Seiji and {Gentile}, Fabrizio and {Gillman}, Steven and {Gould}, Katriona M.~L. and {Gozaliasl}, Ghassem and {Hayward}, Christopher C. and {He}, Qiuhan and {Hemmati}, Shoubaneh and {Hirschmann}, Michaela and {Jahnke}, Knud and {Jin}, Shuowen and {Khostovan}, Ali Ahmad and {Kokorev}, Vasily and {Lambrides}, Erini and {Laigle}, Clotilde and {Larson}, Rebecca L. and {Leung}, Gene C.~K. and {Liu}, Daizhong and {Liaudat}, Tobias and {Long}, Arianna S. and {Magdis}, Georgios and {Mahler}, Guillaume and {Mainieri}, Vincenzo and {Manning}, Sinclaire M. and {Maraston}, Claudia and {Martin}, Crystal L. and {McCleary}, Jacqueline E. and {McKinney}, Jed and {McPartland}, Conor J.~R. and {Mobasher}, Bahram and {Pattnaik}, Rohan and {Renzini}, Alvio and {Rich}, R. Michael and {Sanders}, David B. and {Sattari}, Zahra and {Scognamiglio}, Diana and {Scoville}, Nick and {Sheth}, Kartik and {Shuntov}, Marko and {Sparre}, Martin and {Suzuki}, Tomoko L. and {Talia}, Margherita and {Toft}, Sune and {Trakhtenbrot}, Benny and {Urry}, C. Megan and {Valentino}, Francesco and {Vanderhoof}, Brittany N. and {Vardoulaki}, Eleni and {Weaver}, John R. and {Whitaker}, Katherine E. and {Wilkins}, Stephen M. and {Yang}, Lilan and {Zavala}, Jorge A.},
        title = "{COSMOS-Web: An Overview of the JWST Cosmic Origins Survey}",
      journal = {\apj},
         year = 2023,
        month = sep,
       volume = {954},
       number = {1},
          eid = {31},
        pages = {31},
          doi = {10.3847/1538-4357/acc2bc},
archivePrefix = {arXiv},
       eprint = {2211.07865},
 primaryClass = {astro-ph.GA},
       adsurl = {https://ui.adsabs.harvard.edu/abs/2023ApJ...954...31C}
}

@ARTICLE{cerulo2016,
       author = {{Cerulo}, P. and {Couch}, W.~J. and {Lidman}, C. and {Demarco}, R. and {Huertas-Company}, M. and {Mei}, S. and {S{\'a}nchez-Janssen}, R. and {Barrientos}, L.~F. and {Mu{\~n}oz}, R.~P.},
        title = "{The accelerated build-up of the red sequence in high-redshift galaxy clusters}",
      journal = {\mnras},
         year = 2016,
        month = apr,
       volume = {457},
       number = {2},
        pages = {2209-2235},
          doi = {10.1093/mnras/stw080},
archivePrefix = {arXiv},
       eprint = {1601.07578},
 primaryClass = {astro-ph.GA},
       adsurl = {https://ui.adsabs.harvard.edu/abs/2016MNRAS.457.2209C}
}

@ARTICLE{chung2007,
   author = {{Chung}, A. and {van Gorkom}, J.~H. and {Kenney}, J.~D.~P. and
	{Vollmer}, B.},
    title = "{Virgo Galaxies with Long One-sided H I Tails}",
  journal = {ApJl},
   eprint = {astro-ph/0703338},
     year = 2007,
    month = apr,
   volume = 659,
    pages = {L115-L119},
      doi = {10.1086/518034},
   adsurl = {http://adsabs.harvard.edu/abs/2007ApJ...659L.115C}
}

@ARTICLE{copetti2002,
       author = {{Copetti}, M.~V.~F. and {Writzl}, B.~C.},
        title = "{Study of electron density in planetary nebulae. A comparison of different density indicators}",
      journal = {\aap},
         year = 2002,
        month = jan,
       volume = {382},
        pages = {282-290},
          doi = {10.1051/0004-6361:20011621},
       adsurl = {https://ui.adsabs.harvard.edu/abs/2002A&A...382..282C}
}

@ARTICLE{cortese2007,
       author = {{Cortese}, L. and {Marcillac}, D. and {Richard}, J. and {Bravo-Alfaro}, H. and {Kneib}, J. -P. and {Rieke}, G. and {Covone}, G. and {Egami}, E. and {Rigby}, J. and {Czoske}, O. and {Davies}, J.},
        title = "{The strong transformation of spiral galaxies infalling into massive clusters at z \raisebox{-0.5ex}\textasciitilde 0.2}",
      journal = {MNRAS},
         year = 2007,
        month = mar,
       volume = {376},
       number = {1},
        pages = {157-172},
          doi = {10.1111/j.1365-2966.2006.11369.x},
archivePrefix = {arXiv},
       eprint = {astro-ph/0703012},
 primaryClass = {astro-ph},
       adsurl = {https://ui.adsabs.harvard.edu/abs/2007MNRAS.376..157C}
}

@ARTICLE{cortese2012,
       author = {{Cortese}, L. and {Boissier}, S. and {Boselli}, A. and {Bendo}, G.~J. and {Buat}, V. and {Davies}, J.~I. and {Eales}, S. and {Heinis}, S. and {Isaak}, K.~G. and {Madden}, S.~C.},
        title = "{The GALEX view of the Herschel Reference Survey. Ultraviolet structural properties of nearby galaxies}",
      journal = {A\&A},
         year = 2012,
        month = aug,
       volume = {544},
          eid = {A101},
        pages = {A101},
          doi = {10.1051/0004-6361/201219312},
archivePrefix = {arXiv},
       eprint = {1206.1130},
 primaryClass = {astro-ph.CO},
       adsurl = {https://ui.adsabs.harvard.edu/abs/2012A&A...544A.101C}
}

@ARTICLE{cortese2021,
       author = {{Cortese}, L. and {Catinella}, B. and {Smith}, R.},
        title = "{The Dawes Review 9: The role of cold gas stripping on the star formation quenching of satellite galaxies}",
      journal = {PASA},
         year = 2021,
        month = aug,
       volume = {38},
          eid = {e035},
        pages = {e035},
          doi = {10.1017/pasa.2021.18},
archivePrefix = {arXiv},
       eprint = {2104.02193},
 primaryClass = {astro-ph.GA},
       adsurl = {https://ui.adsabs.harvard.edu/abs/2021PASA...38...35C}
}

@ARTICLE{crossett2025,
       author = {{Crossett}, Jacob P. and {Jaff{\'e}}, Yara L. and {McGee}, Sean L. and {Smith}, Rory and {Bellhouse}, Callum and {Bettoni}, Daniela and {Vulcani}, Benedetta and {Kelkar}, Kshitija and {Louren{\c{c}}o}, Ana C.~C.},
        title = "{Identification of ram pressure stripping features in galaxies using citizen science}",
      journal = {\aap},
         year = 2025,
        month = feb,
       volume = {694},
          eid = {A204},
        pages = {A204},
          doi = {10.1051/0004-6361/202450371},
archivePrefix = {arXiv},
       eprint = {2412.10060},
 primaryClass = {astro-ph.GA},
       adsurl = {https://ui.adsabs.harvard.edu/abs/2025A&A...694A.204C}
}

@ARTICLE{depropris1999,
       author = {{de Propris}, Roberto and {Stanford}, S.~A. and {Eisenhardt}, Peter R. and {Dickinson}, Mark and {Elston}, Richard},
        title = "{The K-Band Luminosity Function in Galaxy Clusters to Z \raisebox{-0.5ex}\textasciitilde1}",
      journal = {\aj},
         year = 1999,
        month = aug,
       volume = {118},
       number = {2},
        pages = {719-729},
          doi = {10.1086/300978},
archivePrefix = {arXiv},
       eprint = {astro-ph/9905137},
 primaryClass = {astro-ph},
       adsurl = {https://ui.adsabs.harvard.edu/abs/1999AJ....118..719D}
}

@ARTICLE{desi2023,
       author = {{DESI Collaboration} and {Adame}, A.~G. and {Aguilar}, J. and {Ahlen}, S. and {Alam}, S. and {Aldering}, G. and {Alexander}, D.~M. and {Alfarsy}, R. and {Allende Prieto}, C. and {Alvarez}, M. and {Alves}, O. and {Anand}, A. and {Andrade-Oliveira}, F. and {Armengaud}, E. and {Asorey}, J. and {Avila}, S. and {Aviles}, A. and {Bailey}, S. and {Balaguera-Antol{\'\i}nez}, A. and {Ballester}, O. and {Baltay}, C. and {Bault}, A. and {Bautista}, J. and {Behera}, J. and {Beltran}, S.~F. and {BenZvi}, S. and {Beraldo e Silva}, L. and {Bermejo-Climent}, J.~R. and {Berti}, A. and {Besuner}, R. and {Beutler}, F. and {Bianchi}, D. and {Blake}, C. and {Blum}, R. and {Bolton}, A.~S. and {Brieden}, S. and {Brodzeller}, A. and {Brooks}, D. and {Brown}, Z. and {Buckley-Geer}, E. and {Burtin}, E. and {Cabayol-Garcia}, L. and {Cai}, Z. and {Canning}, R. and {Cardiel-Sas}, L. and {Carnero Rosell}, A. and {Castander}, F.~J. and {Cervantes-Cota}, J.~L. and {Chabanier}, S. and {Chaussidon}, E. and {Chaves-Montero}, J. and {Chen}, S. and {Chen}, X. and {Chuang}, C. and {Claybaugh}, T. and {Cole}, S. and {Cooper}, A.~P. and {Cuceu}, A. and {Davis}, T.~M. and {Dawson}, K. and {de Belsunce}, R. and {de la Cruz}, R. and {de la Macorra}, A. and {Della Costa}, J. and {de Mattia}, A. and {Demina}, R. and {Demirbozan}, U. and {DeRose}, J. and {Dey}, A. and {Dey}, B. and {Dhungana}, G. and {Ding}, J. and {Ding}, Z. and {Doel}, P. and {Doshi}, R. and {Douglass}, K. and {Edge}, A. and {Eftekharzadeh}, S. and {Eisenstein}, D.~J. and {Elliott}, A. and {Ereza}, J. and {Escoffier}, S. and {Fagrelius}, P. and {Fan}, X. and {Fanning}, K. and {Fawcett}, V.~A. and {Ferraro}, S. and {Flaugher}, B. and {Font-Ribera}, A. and {Forero-Romero}, J.~E. and {Forero-S{\'a}nchez}, D. and {Frenk}, C.~S. and {G{\"a}nsicke}, B.~T. and {Garc{\'\i}a}, L. {\'A}. and {Garc{\'\i}a-Bellido}, J. and {Garcia-Quintero}, C. and {Garrison}, L.~H. and {Gil-Mar{\'\i}n}, H. and {Golden-Marx}, J. and {Gontcho A Gontcho}, S. and {Gonzalez-Morales}, A.~X. and {Gonzalez-Perez}, V. and {Gordon}, C. and {Graur}, O. and {Green}, D. and {Gruen}, D. and {Guy}, J. and {Hadzhiyska}, B. and {Hahn}, C. and {Han}, J.~J. and {Hanif}, M.~M.~S. and {Herrera-Alcantar}, H.~K. and {Honscheid}, K. and {Hou}, J. and {Howlett}, C. and {Huterer}, D. and {Ir{\v{s}}i{\v{c}}}, V. and {Ishak}, M. and {Jacques}, A. and {Jana}, A. and {Jiang}, L. and {Jimenez}, J. and {Jing}, Y.~P. and {Joudaki}, S. and {Joyce}, R. and {Jullo}, E. and {Juneau}, S. and {Kara{\c{c}}ayl{\i}}, N.~G. and {Karim}, T. and {Kehoe}, R. and {Kent}, S. and {Khederlarian}, A. and {Kim}, S. and {Kirkby}, D. and {Kisner}, T. and {Kitaura}, F. and {Kizhuprakkat}, N. and {Kneib}, J. and {Koposov}, S.~E. and {Kov{\'a}cs}, A. and {Kremin}, A. and {Krolewski}, A. and {L'Huillier}, B. and {Lahav}, O. and {Lambert}, A. and {Lamman}, C. and {Lan}, T. -W. and {Landriau}, M. and {Lang}, D. and {Lange}, J.~U. and {Lasker}, J. and {Leauthaud}, A. and {Le Guillou}, L. and {Levi}, M.~E. and {Li}, T.~S. and {Linder}, E. and {Lyons}, A. and {Magneville}, C. and {Manera}, M. and {Manser}, C.~J. and {Margala}, D. and {Martini}, P. and {McDonald}, P. and {Medina}, G.~E. and {Medina-Varela}, L. and {Meisner}, A. and {Mena-Fern{\'a}ndez}, J. and {Meneses-Rizo}, J. and {Mezcua}, M. and {Miquel}, R. and {Montero-Camacho}, P. and {Moon}, J. and {Moore}, S. and {Moustakas}, J. and {Mueller}, E. and {Mundet}, J. and {Mu{\~n}oz-Guti{\'e}rrez}, A. and {Myers}, A.~D. and {Nadathur}, S. and {Napolitano}, L. and {Neveux}, R. and {Newman}, J.~A. and {Nie}, J. and {Nikutta}, R. and {Niz}, G. and {Norberg}, P. and {Noriega}, H.~E. and {Paillas}, E. and {Palanque-Delabrouille}, N. and {Palmese}, A. and {Pan}, Z. and {Parkinson}, D. and {Penmetsa}, S. and {Percival}, W.~J. and {P{\'e}rez-Fern{\'a}ndez}, A. and {P{\'e}rez-R{\`a}fols}, I. and {Pieri}, M. and {Poppett}, C. and {Porredon}, A. and {Pothier}, S.},
        title = "{The Early Data Release of the Dark Energy Spectroscopic Instrument}",
      journal = {\aj},
         year = 2024,
        month = aug,
       volume = {168},
       number = {2},
          eid = {58},
        pages = {58},
          doi = {10.3847/1538-3881/ad3217},
archivePrefix = {arXiv},
       eprint = {2306.06308},
 primaryClass = {astro-ph.CO},
       adsurl = {https://ui.adsabs.harvard.edu/abs/2024AJ....168...58D}
}

@ARTICLE{deg2023,
       author = {{Deg}, N. and {Palleske}, R. and {Spekkens}, K. and {Wang}, J. and {Jarrett}, T. and {English}, J. and {Lin}, X. and {Yeung}, J. and {Mould}, J.~R. and {Catinella}, B. and {D{\'e}nes}, H. and {Elagali}, A. and {For}, B.-Q. and {Kamphuis}, P. and {Koribalski}, B.~S. and {Lee-Waddell}, K. and {Murugeshan}, C. and {Oh}, S. and {Rhee}, J. and {Serra}, P. and {Westmeier}, T. and {Wong}, O.~I. and {Bekki}, K. and {Bosma}, A. and {Carignan}, C. and {Holwerda}, B.~W. and {Yu}, N.},
        title = "{WALLABY pilot survey: the potential polar ring galaxies NGC 4632 and NGC 6156}",
      journal = {\mnras},
         year = 2023,
        month = nov,
       volume = {525},
       number = {3},
        pages = {4663-4684},
          doi = {10.1093/mnras/stad2312},
archivePrefix = {arXiv},
       eprint = {2309.05841},
 primaryClass = {astro-ph.GA},
       adsurl = {https://ui.adsabs.harvard.edu/abs/2023MNRAS.525.4663D}
}

@ARTICLE{dressler1980,
   author = {{Dressler}, A.},
    title = "{Galaxy morphology in rich clusters - Implications for the formation and evolution of galaxies}",
  journal = {ApJ},
     year = 1980,
    month = mar,
   volume = 236,
    pages = {351-365},
      doi = {10.1086/157753},
   adsurl = {http://adsabs.harvard.edu/abs/1980ApJ...236..351D}
}

@ARTICLE{durret2021,
       author = {{Durret}, F. and {Chiche}, S. and {Lobo}, C. and {Jauzac}, M.},
        title = "{Jellyfish galaxy candidates in MACS J0717.5+3745 and 39 other clusters of the DAFT/FADA and CLASH surveys}",
      journal = {A\&A},
         year = 2021,
        month = apr,
       volume = {648},
          eid = {A63},
        pages = {A63},
          doi = {10.1051/0004-6361/202039770},
archivePrefix = {arXiv},
       eprint = {2102.02595},
 primaryClass = {astro-ph.GA},
       adsurl = {https://ui.adsabs.harvard.edu/abs/2021A&A...648A..63D}
}

@ARTICLE{ebeling2014,
       author = {{Ebeling}, H. and {Stephenson}, L.~N. and {Edge}, A.~C.},
        title = "{Jellyfish: Evidence of Extreme Ram-pressure Stripping in Massive Galaxy Clusters}",
      journal = {ApJL},
         year = 2014,
        month = feb,
       volume = {781},
       number = {2},
          eid = {L40},
        pages = {L40},
          doi = {10.1088/2041-8205/781/2/L40},
archivePrefix = {arXiv},
       eprint = {1312.6135},
 primaryClass = {astro-ph.GA},
       adsurl = {https://ui.adsabs.harvard.edu/abs/2014ApJ...781L..40E}
}

@ARTICLE{edler2023,
       author = {{Edler}, H.~W. and {de Gasperin}, F. and {Shimwell}, T.~W. and {Hardcastle}, M.~J. and {Boselli}, A. and {Heesen}, V. and {McCall}, H. and {Bomans}, D.~J. and {Br{\"u}ggen}, M. and {Bulbul}, E. and {Chy{\.z}y}, K.~T. and {Ignesti}, A. and {Merloni}, A. and {Pacaud}, F. and {Reiprich}, T.~H. and {Roberts}, I.~D. and {Rottgering}, H.~J.~A. and {van Weeren}, R.~J.},
        title = "{VICTORIA project: The LOFAR HBA Virgo Cluster Survey}",
      journal = {\aap},
         year = 2023,
        month = aug,
       volume = {676},
          eid = {A24},
        pages = {A24},
          doi = {10.1051/0004-6361/202346458},
archivePrefix = {arXiv},
       eprint = {2306.04513},
 primaryClass = {astro-ph.GA},
       adsurl = {https://ui.adsabs.harvard.edu/abs/2023A&A...676A..24E}
}

@ARTICLE{ferland2017,
       author = {{Ferland}, G.~J. and {Chatzikos}, M. and {Guzm{\'a}n}, F. and {Lykins}, M.~L. and {van Hoof}, P.~A.~M. and {Williams}, R.~J.~R. and {Abel}, N.~P. and {Badnell}, N.~R. and {Keenan}, F.~P. and {Porter}, R.~L. and {Stancil}, P.~C.},
        title = "{The 2017 Release Cloudy}",
      journal = {\rmxaa},
         year = 2017,
        month = oct,
       volume = {53},
        pages = {385-438},
          doi = {10.48550/arXiv.1705.10877},
archivePrefix = {arXiv},
       eprint = {1705.10877},
 primaryClass = {astro-ph.GA},
       adsurl = {https://ui.adsabs.harvard.edu/abs/2017RMxAA..53..385F}
}

@ARTICLE{finn2018,
       author = {{Finn}, Rose A. and {Desai}, Vandana and {Rudnick}, Gregory and
         {Balogh}, Michael and {Haynes}, Martha P. and {Jablonka}, Pascale and
         {Koopmann}, Rebecca A. and {Moustakas}, John and {Peng}, Chien Y. and
         {Poggianti}, Bianca and {Rines}, Kenneth and {Zaritsky}, Dennis},
        title = "{The Local Cluster Survey. I. Evidence of Outside-in Quenching in Dense Environments}",
      journal = {ApJ},
         year = 2018,
        month = aug,
       volume = {862},
       number = {2},
          eid = {149},
        pages = {149},
          doi = {10.3847/1538-4357/aac32a},
archivePrefix = {arXiv},
       eprint = {1807.03388},
 primaryClass = {astro-ph.GA},
       adsurl = {https://ui.adsabs.harvard.edu/abs/2018ApJ...862..149F}
}

@ARTICLE{finkleman2011,
       author = {{Finkelman}, Ido and {Moiseev}, Alexei and {Brosch}, Noah and {Katkov}, Ivan},
        title = "{Hoag's Object: evidence for cold accretion on to an elliptical galaxy}",
      journal = {\mnras},
         year = 2011,
        month = dec,
       volume = {418},
       number = {3},
        pages = {1834-1849},
          doi = {10.1111/j.1365-2966.2011.19601.x},
archivePrefix = {arXiv},
       eprint = {1108.3079},
 primaryClass = {astro-ph.CO},
       adsurl = {https://ui.adsabs.harvard.edu/abs/2011MNRAS.418.1834F}
}

@article{finn2025,
   title={Virgo Filaments. V. Disrupting the Baryon Cycle in the NGC 5364 Galaxy Group},
   volume={985},
   ISSN={1538-4357},
   url={http://dx.doi.org/10.3847/1538-4357/adc566},
   DOI={10.3847/1538-4357/adc566},
   number={1},
   journal={The Astrophysical Journal},
   publisher={American Astronomical Society},
   author={Finn, Rose A. and Rudnick, Gregory and Jablonka, Pascale and Ramatsoku, Mpati and Nagaraj, Gautam and Vulcani, Benedetta and Koopmann, Rebecca A. and Fossati, Matteo and Agostino, James and Bahé, Yannick and Garcia-Burillo, Santiago and Castignani, Gianluca and Combes, Francoise and Conger, Kim and De Lucia, Gabriella and Desai, Vandana and Moustakas, John and Norman, Dara and Sperone-Longin, Damien and Townsend, Melinda and Xie, Lizhi and Zakharova, Daria and Zaritsky, Dennis},
   year={2025},
   month=may, pages={81} }

@article{fogarty2011,
   title={SWIFT observations of the Arp 147 ring galaxy system: Arp147},
   volume={417},
   ISSN={0035-8711},
   url={http://dx.doi.org/10.1111/j.1365-2966.2011.19066.x},
   DOI={10.1111/j.1365-2966.2011.19066.x},
   number={2},
   journal={Monthly Notices of the Royal Astronomical Society},
   publisher={Oxford University Press (OUP)},
   author={Fogarty, L. and Thatte, N. and Tecza, M. and Clarke, F. and Goodsall, T. and Houghton, R. and Salter, G. and Davies, R. L. and Kassin, S. A.},
   year={2011},
   month=sep, pages={835–844} }

@ARTICLE{foreman-mackey2013,
   author = {{Foreman-Mackey}, D. and {Hogg}, D.~W. and {Lang}, D. and {Goodman}, J.
	},
    title = "{emcee: The MCMC Hammer}",
  journal = {PASP},
archivePrefix = "arXiv",
   eprint = {1202.3665},
 primaryClass = "astro-ph.IM",
     year = 2013,
    month = mar,
   volume = 125,
    pages = {306},
      doi = {10.1086/670067},
   adsurl = {http://adsabs.harvard.edu/abs/2013PASP..125..306F}
}

@ARTICLE{franco2025,
       author = {{Franco}, Maximilien and {Casey}, Caitlin M. and {Koekemoer}, Anton M. and {Liu}, Daizhong and {Bagley}, Micaela B. and {McCracken}, Henry Joy and {Kartaltepe}, Jeyhan S. and {Akins}, Hollis B. and {Ilbert}, Olivier and {Shuntov}, Marko and {Harish}, Santosh and {Robertson}, Brant E. and {Arango-Toro}, Rafael C. and {Battisti}, Andrew J. and {Chartab}, Nima and {Drakos}, Nicole E. and {Faisst}, Andreas L. and {Flayhart}, Carter and {Gozaliasl}, Ghassem and {Hirschmann}, Michaela and {Massey}, Richard and {Rhodes}, Jason and {Sattari}, Zahra and {Scognamiglio}, Diana and {Weaver}, John R. and {Yang}, Lilan and {Zavala}, Jorge A. and {Berman}, Edward M. and {Gentile}, Fabrizio and {Gillman}, Steven and {Long}, Arianna S. and {Magdis}, Georgios and {McCleary}, Jacqueline E. and {McKinney}, Jed and {Mobasher}, Bahram and {Paquereau}, Louise and {Rest}, Armin and {Sanders}, David B. and {Toft}, Sune and {Yu}, Si-Yue},
        title = "{COSMOS-Web: Comprehensive Data Reduction for Wide-Area JWST NIRCam Imaging}",
      journal = {arXiv e-prints},
         year = 2025,
        month = jun,
          eid = {arXiv:2506.03256},
        pages = {arXiv:2506.03256},
          doi = {10.48550/arXiv.2506.03256},
archivePrefix = {arXiv},
       eprint = {2506.03256},
 primaryClass = {astro-ph.IM},
       adsurl = {https://ui.adsabs.harvard.edu/abs/2025arXiv250603256F}
}

@ARTICLE{fumagalli2011,
       author = {{Fumagalli}, Mattia and {Gavazzi}, G. and {Scaramella}, R. and {Franzetti}, P.},
        title = "{Constraining the ages of the fireballs in the wake of the dIrr galaxy VCC 1217/IC 3418}",
      journal = {\aap},
         year = 2011,
        month = apr,
       volume = {528},
          eid = {A46},
        pages = {A46},
          doi = {10.1051/0004-6361/201015463},
archivePrefix = {arXiv},
       eprint = {1011.1665},
 primaryClass = {astro-ph.CO},
       adsurl = {https://ui.adsabs.harvard.edu/abs/2011A&A...528A..46F}
}

@ARTICLE{gavazzi1978,
       author = {{Gavazzi}, G.},
        title = "{A Westerbork survey of clusters of galaxies. VII. Peculiar radio sources in the central region of Abell 1367.}",
      journal = {A\&A},
         year = 1978,
        month = oct,
       volume = {69},
        pages = {355-361},
       adsurl = {https://ui.adsabs.harvard.edu/abs/1978A&A....69..355G}
}

@ARTICLE{gavazzi1987_a1367,
       author = {{Gavazzi}, G. and {Jaffe}, W.},
        title = "{50 KPC radio trails behind irregular galaxies in A 1367.}",
      journal = {A\&A},
         year = 1987,
        month = nov,
       volume = {186},
        pages = {L1-L2},
       adsurl = {https://ui.adsabs.harvard.edu/abs/1987A&A...186L...1G}
}

@ARTICLE{gavazzi2001,
       author = {{Gavazzi}, G. and {Boselli}, A. and {Mayer}, L. and {Iglesias-Paramo}, J. and {V{\'\i}lchez}, J.~M. and {Carrasco}, L.},
        title = "{75 Kiloparsec Trails of Ionized Gas behind Two Irregular Galaxies in A1367}",
      journal = {ApJL},
         year = 2001,
        month = dec,
       volume = {563},
       number = {1},
        pages = {L23-L26},
          doi = {10.1086/338389},
archivePrefix = {arXiv},
       eprint = {astro-ph/0111085},
 primaryClass = {astro-ph},
       adsurl = {https://ui.adsabs.harvard.edu/abs/2001ApJ...563L..23G}
}

@ARTICLE{gomez-gonzalez2024,
       author = {{G{\'o}mez-Gonz{\'a}lez}, V.~M.~A. and {Mayya}, Y.~D. and {Zaragoza-Cardiel}, J. and {Bruzual}, G. and {Charlot}, S. and {Ramos-Larios}, G. and {Oskinova}, L.~M. and {Sander}, A.~A.~C. and {Serantes}, S. Reyero},
        title = "{Chemical abundances and ionizing mechanisms in the star-forming double-ring of AM 0644-741 using MUSE data}",
      journal = {\mnras},
         year = 2024,
        month = apr,
       volume = {529},
       number = {4},
        pages = {4369-4386},
          doi = {10.1093/mnras/stae570},
archivePrefix = {arXiv},
       eprint = {2402.13230},
 primaryClass = {astro-ph.GA},
       adsurl = {https://ui.adsabs.harvard.edu/abs/2024MNRAS.529.4369G}
}

@ARTICLE{gullieuszik2017,
       author = {{Gullieuszik}, Marco and {Poggianti}, Bianca M. and {Moretti}, Alessia and {Fritz}, Jacopo and {Jaff{\'e}}, Yara L. and {Hau}, George and {Bischko}, Jan C. and {Bellhouse}, Callum and {Bettoni}, Daniela and {Fasano}, Giovanni and {Vulcani}, Benedetta and {D'Onofrio}, Mauro and {Biviano}, Andrea},
        title = "{GASP. IV. A Muse View of Extreme Ram-pressure-stripping in the Plane of the Sky: The Case of Jellyfish Galaxy JO204}",
      journal = {\apj},
         year = 2017,
        month = sep,
       volume = {846},
       number = {1},
          eid = {27},
        pages = {27},
          doi = {10.3847/1538-4357/aa8322},
archivePrefix = {arXiv},
       eprint = {1708.09035},
 primaryClass = {astro-ph.GA},
       adsurl = {https://ui.adsabs.harvard.edu/abs/2017ApJ...846...27G}
}

@ARTICLE{gunn1972,
   author = {{Gunn}, J.~E. and {Gott}, III, J.~R.},
    title = "{On the Infall of Matter Into Clusters of Galaxies and Some Effects on Their Evolution}",
  journal = {ApJ},
     year = 1972,
    month = aug,
   volume = 176,
    pages = {1},
      doi = {10.1086/151605},
   adsurl = {http://adsabs.harvard.edu/abs/1972ApJ...176....1G}
}

@ARTICLE{haines2015,
   author = {{Haines}, C.~P. and {Pereira}, M.~J. and {Smith}, G.~P. and
	{Egami}, E. and {Babul}, A. and {Finoguenov}, A. and {Ziparo}, F. and
	{McGee}, S.~L. and {Rawle}, T.~D. and {Okabe}, N. and {Moran}, S.~M.
	},
    title = "{LoCuSS: The Slow Quenching of Star Formation in Cluster Galaxies and the Need for Pre-processing}",
  journal = {ApJ},
archivePrefix = "arXiv",
   eprint = {1504.05604},
     year = 2015,
    month = jun,
   volume = 806,
      eid = {101},
    pages = {101},
      doi = {10.1088/0004-637X/806/1/101},
   adsurl = {http://adsabs.harvard.edu/abs/2015ApJ...806..101H}
}

@Article{harris2020,
 title         = {Array programming with {NumPy}},
 author        = {Charles R. Harris and K. Jarrod Millman and St{\'{e}}fan J.
                 van der Walt and Ralf Gommers and Pauli Virtanen and David
                 Cournapeau and Eric Wieser and Julian Taylor and Sebastian
                 Berg and Nathaniel J. Smith and Robert Kern and Matti Picus
                 and Stephan Hoyer and Marten H. van Kerkwijk and Matthew
                 Brett and Allan Haldane and Jaime Fern{\'{a}}ndez del
                 R{\'{i}}o and Mark Wiebe and Pearu Peterson and Pierre
                 G{\'{e}}rard-Marchant and Kevin Sheppard and Tyler Reddy and
                 Warren Weckesser and Hameer Abbasi and Christoph Gohlke and
                 Travis E. Oliphant},
 year          = {2020},
 month         = sep,
 journal       = {Nature},
 volume        = {585},
 number        = {7825},
 pages         = {357--362},
 doi           = {10.1038/s41586-020-2649-2},
 publisher     = {Springer Science and Business Media {LLC}},
 url           = {https://doi.org/10.1038/s41586-020-2649-2}
}

@ARTICLE{hamadouche2024,
       author = {{Hamadouche}, M.~L. and {McLure}, R.~J. and {Carnall}, A. and {McLeod}, D.~J. and {Dunlop}, J.~S. and {Whitaker}, K. and {Donnan}, C.~T. and {Begley}, R. and {Stanton}, T.~M. and {Almaini}, O. and {Aird}, J. and {Cullen}, F. and {Cutler}, S. and {Koekemoer}, A.~M.},
        title = "{JWST PRIMER: strong evidence for the environmental quenching of low-mass galaxies out to $\mathbf{\textit{z} \simeq 2}$}",
      journal = {arXiv e-prints},
         year = 2024,
        month = dec,
          eid = {arXiv:2412.09592},
        pages = {arXiv:2412.09592},
          doi = {10.48550/arXiv.2412.09592},
archivePrefix = {arXiv},
       eprint = {2412.09592},
 primaryClass = {astro-ph.GA},
       adsurl = {https://ui.adsabs.harvard.edu/abs/2024arXiv241209592H}
}

@ARTICLE{hester2010,
       author = {{Hester}, Janice A. and {Seibert}, Mark and {Neill}, James D. and {Wyder}, Ted K. and {Gil de Paz}, Armando and {Madore}, Barry F. and {Martin}, D. Christopher and {Schiminovich}, David and {Rich}, R. Michael},
        title = "{IC 3418: Star Formation in a Turbulent Wake}",
      journal = {\apjl},
         year = 2010,
        month = jun,
       volume = {716},
       number = {1},
        pages = {L14-L18},
          doi = {10.1088/2041-8205/716/1/L14},
archivePrefix = {arXiv},
       eprint = {1006.5746},
 primaryClass = {astro-ph.GA},
       adsurl = {https://ui.adsabs.harvard.edu/abs/2010ApJ...716L..14H}
}

@ARTICLE{higdon1995,
       author = {{Higdon}, James L.},
        title = "{Wheels of Fire. I. Massive Star Formation in the Cartwheel Ring Galaxy}",
      journal = {\apj},
         year = 1995,
        month = dec,
       volume = {455},
        pages = {524},
          doi = {10.1086/176602},
       adsurl = {https://ui.adsabs.harvard.edu/abs/1995ApJ...455..524H}
}

@ARTICLE{higdon1996,
       author = {{Higdon}, James L.},
        title = "{Wheels of Fire. II. Neutral Hydrogen in the Cartwheel Ring Galaxy}",
      journal = {\apj},
         year = 1996,
        month = aug,
       volume = {467},
        pages = {241},
          doi = {10.1086/177599},
       adsurl = {https://ui.adsabs.harvard.edu/abs/1996ApJ...467..241H}
}

@ARTICLE{higdon2011,
       author = {{Higdon}, James L. and {Higdon}, Sarah J.~U. and {Rand}, Richard J.},
        title = "{Wheels of Fire. IV. Star Formation and the Neutral Interstellar Medium in the Ring Galaxy AM0644-741}",
      journal = {\apj},
         year = 2011,
        month = oct,
       volume = {739},
       number = {2},
          eid = {97},
        pages = {97},
          doi = {10.1088/0004-637X/739/2/97},
archivePrefix = {arXiv},
       eprint = {1107.5570},
 primaryClass = {astro-ph.CO},
       adsurl = {https://ui.adsabs.harvard.edu/abs/2011ApJ...739...97H}
}

@ARTICLE{hosseinzadeh2023,
       author = {{Hosseinzadeh}, Griffin and {Sand}, David J. and {Jencson}, Jacob E. and {Andrews}, Jennifer E. and {Shivaei}, Irene and {Bostroem}, K. Azalee and {Valenti}, Stefano and {Szalai}, Tam{\'a}s and {Burke}, Jamison and {Howell}, D. Andrew and {McCully}, Curtis and {Newsome}, Megan and {Gonzalez}, Estefania Padilla and {Pellegrino}, Craig and {Terreran}, Giacomo},
        title = "{JWST Imaging of the Cartwheel Galaxy Reveals Dust Associated with SN 2021afdx}",
      journal = {\apjl},
         year = 2023,
        month = jan,
       volume = {942},
       number = {1},
          eid = {L18},
        pages = {L18},
          doi = {10.3847/2041-8213/aca64e},
archivePrefix = {arXiv},
       eprint = {2210.06499},
 primaryClass = {astro-ph.HE},
       adsurl = {https://ui.adsabs.harvard.edu/abs/2023ApJ...942L..18H}
}

@ARTICLE{hu2024,
       author = {{Hu}, Dan and {Zaja{\v{c}}ek}, Michal and {Werner}, Norbert and {Grossov{\'a}}, Romana and {J{\'a}chym}, Pavel and {Roberts}, Ian D. and {Ignesti}, Alessandro and {Kenney}, Jeffrey D.~P. and {Pl{\v{s}}ek}, Tom{\'a}{\v{s}} and {Breuer}, Jean-Paul and {Shimwell}, Timothy and {Tasse}, Cyril and {Zhu}, Zhenghao and {Wu}, Linhui},
        title = "{Ram-pressure stripped radio tail and two ULXs in the spiral galaxy HCG 97b}",
      journal = {\mnras},
         year = 2024,
        month = jan,
       volume = {527},
       number = {1},
        pages = {1062-1080},
          doi = {10.1093/mnras/stad3219},
archivePrefix = {arXiv},
       eprint = {2304.13066},
 primaryClass = {astro-ph.GA},
       adsurl = {https://ui.adsabs.harvard.edu/abs/2024MNRAS.527.1062H}
}

@Article{hunter2007,
  Author    = {Hunter, J. D.},
  Title     = {Matplotlib: A 2D graphics environment},
  Journal   = {Computing In Science \& Engineering},
  Volume    = {9},
  Number    = {3},
  Pages     = {90--95},
  publisher = {IEEE COMPUTER SOC},
  doi = {10.1109/MCSE.2007.55},
  year      = 2007
}

@ARTICLE{ignesti2023_a2255,
       author = {{Ignesti}, A. and {Vulcani}, B. and {Botteon}, A. and {Poggianti}, B. and {Giunchi}, E. and {Smith}, R. and {Brunetti}, G. and {Roberts}, I.~D. and {van Weeren}, R.~J. and {Rajpurohit}, K.},
        title = "{Radio continuum tails in ram pressure-stripped spiral galaxies: Experimenting with a semi-empirical model in Abell 2255}",
      journal = {\aap},
         year = 2023,
        month = jul,
       volume = {675},
          eid = {A118},
        pages = {A118},
          doi = {10.1051/0004-6361/202346517},
archivePrefix = {arXiv},
       eprint = {2305.19941},
 primaryClass = {astro-ph.GA},
       adsurl = {https://ui.adsabs.harvard.edu/abs/2023A&A...675A.118I}
}

@ARTICLE{jachym2014,
       author = {{J{\'a}chym}, Pavel and {Combes}, Fran{\c{c}}oise and {Cortese}, Luca and
         {Sun}, Ming and {Kenney}, Jeffrey D.~P.},
        title = "{Abundant Molecular Gas and Inefficient Star Formation in Intracluster Regions: Ram Pressure Stripped Tail of the Norma Galaxy ESO137-001}",
      journal = {ApJ},
         year = 2014,
        month = sep,
       volume = {792},
       number = {1},
          eid = {11},
        pages = {11},
          doi = {10.1088/0004-637X/792/1/11},
archivePrefix = {arXiv},
       eprint = {1403.2328},
 primaryClass = {astro-ph.GA},
       adsurl = {https://ui.adsabs.harvard.edu/abs/2014ApJ...792...11J}
}

@ARTICLE{jachym2019,
       author = {{J{\'a}chym}, Pavel and {Kenney}, Jeffrey D.~P. and {Sun}, Ming and
         {Combes}, Fran{\c{c}}oise and {Cortese}, Luca and {Scott}, Tom C. and
         {Sivanandam}, Suresh and {Brinks}, Elias and {Roediger}, Elke and
         {Palou{\v{s}}}, Jan and {Fumagalli}, Michele},
        title = "{ALMA Unveils Widespread Molecular Gas Clumps in the Ram Pressure Stripped Tail of the Norma Jellyfish Galaxy}",
      journal = {ApJ},
         year = 2019,
        month = oct,
       volume = {883},
       number = {2},
          eid = {145},
        pages = {145},
          doi = {10.3847/1538-4357/ab3e6c},
archivePrefix = {arXiv},
       eprint = {1905.13249},
 primaryClass = {astro-ph.GA},
       adsurl = {https://ui.adsabs.harvard.edu/abs/2019ApJ...883..145J}
}

@ARTICLE{jaffe2018,
   author = {{Jaff{\'e}}, Y.~L. and {Poggianti}, B.~M. and {Moretti}, A. and
	{Gullieuszik}, M. and {Smith}, R. and {Vulcani}, B. and {Fasano}, G. and
	{Fritz}, J. and {Tonnesen}, S. and {Bettoni}, D. and {Hau}, G. and
	{Biviano}, A. and {Bellhouse}, C. and {McGee}, S.},
    title = "{GASP IX. Jellyfish galaxies in phase-space: an orbital study of intense ram-pressure stripping in clusters}",
  journal = {MNRAS},
     year = 2018,
    month = feb,
      doi = {10.1093/mnras/sty500},
   adsurl = {http://adsabs.harvard.edu/abs/2018MNRAS.tmp..484J}
}

@ARTICLE{jimenez-teja2019,
       author = {{Jim{\'e}nez-Teja}, Y. and {Dupke}, R.~A. and {Lopes de Oliveira}, R. and {Xavier}, H.~S. and {Coelho}, P.~R.~T. and {Chies-Santos}, A.~L. and {L{\'o}pez-Sanjuan}, C. and {Alvarez-Candal}, A. and {Costa-Duarte}, M.~V. and {Telles}, E. and {Hernandez-Jimenez}, J.~A. and {Ben{\'\i}tez}, N. and {Alcaniz}, J. and {Cenarro}, J. and {Crist{\'o}bal-Hornillos}, D. and {Ederoclite}, A. and {Mar{\'\i}n-Franch}, A. and {Mendes de Oliveira}, C. and {Moles}, M. and {Sodr{\'e}}, L. and {Varela}, J. and {V{\'a}zquez Rami{\'o}}, H.},
        title = "{J-PLUS: Analysis of the intracluster light in the Coma cluster}",
      journal = {\aap},
         year = 2019,
        month = feb,
       volume = {622},
          eid = {A183},
        pages = {A183},
          doi = {10.1051/0004-6361/201833547},
archivePrefix = {arXiv},
       eprint = {1810.01424},
 primaryClass = {astro-ph.GA},
       adsurl = {https://ui.adsabs.harvard.edu/abs/2019A&A...622A.183J}
}

@ARTICLE{joo2023,
       author = {{Joo}, Hyungjin and {Jee}, M. James},
        title = "{Intracluster light is already abundant at redshift beyond unity}",
      journal = {\nat},
         year = 2023,
        month = jan,
       volume = {613},
       number = {7942},
        pages = {37-41},
          doi = {10.1038/s41586-022-05396-4},
archivePrefix = {arXiv},
       eprint = {2301.01523},
 primaryClass = {astro-ph.GA},
       adsurl = {https://ui.adsabs.harvard.edu/abs/2023Natur.613...37J}
}

@ARTICLE{junais2021,
       author = {{Junais} and {Boissier}, S. and {Boselli}, A. and {Boquien}, M. and {Longobardi}, A. and {Roehlly}, Y. and {Amram}, P. and {Fossati}, M. and {Cuillandre}, J. -C. and {Gwyn}, S. and {Ferrarese}, L. and {C{\^o}t{\'e}}, P. and {Roediger}, J. and {Lim}, S. and {Peng}, E.~W. and {Hensler}, G. and {Trinchieri}, G. and {Koda}, J. and {Prantzos}, N.},
        title = "{A Virgo Environmental Survey Tracing Ionised Gas Emission (VESTIGE). X. Formation of a red ultra-diffuse galaxy and an almost dark galaxy during a ram-pressure stripping event}",
      journal = {\aap},
         year = 2021,
        month = jun,
       volume = {650},
          eid = {A99},
        pages = {A99},
          doi = {10.1051/0004-6361/202040185},
archivePrefix = {arXiv},
       eprint = {2104.02492},
 primaryClass = {astro-ph.GA},
       adsurl = {https://ui.adsabs.harvard.edu/abs/2021A&A...650A..99J}
}

@ARTICLE{katkov2022,
       author = {{Katkov}, Ivan Yu. and {Kniazev}, Alexei Yu. and {Sil'chenko}, Olga K. and {Gasymov}, Damir},
        title = "{Star formation in outer rings of S0 galaxies. IV. NGC 254: A double-ringed S0 with gas counter-rotation}",
      journal = {\aap},
         year = 2022,
        month = feb,
       volume = {658},
          eid = {A154},
        pages = {A154},
          doi = {10.1051/0004-6361/202141934},
archivePrefix = {arXiv},
       eprint = {2112.03289},
 primaryClass = {astro-ph.GA},
       adsurl = {https://ui.adsabs.harvard.edu/abs/2022A&A...658A.154K}
}

@ARTICLE{khostovan2025,
       author = {{Khostovan}, Ali Ahmad and {Kartaltepe}, Jeyhan S. and {Salvato}, Mara and {Ilbert}, Olivier and {Casey}, Caitlin M. and {Algera}, Hiddo and {Antwi-Danso}, Jacqueline and {Battisti}, Andrew and {Brinch}, Malte and {Brusa}, Marcella and {Calabro}, Antonello and {Capak}, Peter L. and {Chartab}, Nima and {Cooper}, Olivia R. and {Cox}, Isa G. and {Darvish}, Behnam and {Drakos}, Nicole E. and {Faisst}, Andreas L. and {George}, Matthew R. and {Gozaliasl}, Ghassem and {Harish}, Santosh and {Hasinger}, Gunther and {Hatamnia}, Hossein and {Iovino}, Angela and {Jin}, Shuowen and {Kashino}, Daichi and {Koekemoer}, Anton M. and {Laishram}, Ronaldo and {Lee}, Khee-Gan and {Lertprasertpong}, Jitrapon and {Lilly}, Simon J. and {Masters}, Daniel C. and {Mobasher}, Bahram and {Nagao}, Tohru and {Onodera}, Masato and {Peng}, Yingjie and {Sanders}, David B. and {Sanders}, Ryan L. and {Sattari}, Zahra and {Scoville}, Nick and {Shah}, Ekta A. and {Silverman}, John D. and {Suzuki}, Nao and {Tanaka}, Masayuki and {Toft}, Sune and {Trakhtenbrot}, Benny and {Trump}, Jonathan R. and {Vaccari}, Mattia and {Valentino}, Francesco and {Vanderhoof}, Brittany N. and {Weaver}, John R. and {Yun}, Min S. and {Zavala}, Jorge A.},
        title = "{COSMOS Spectroscopic Redshift Compilation (First Data Release): 165k Redshifts Encompassing Two Decades of Spectroscopy}",
      journal = {arXiv e-prints},
         year = 2025,
        month = feb,
          eid = {arXiv:2503.00120},
        pages = {arXiv:2503.00120},
          doi = {10.48550/arXiv.2503.00120},
archivePrefix = {arXiv},
       eprint = {2503.00120},
 primaryClass = {astro-ph.GA},
       adsurl = {https://ui.adsabs.harvard.edu/abs/2025arXiv250300120K}
}

@ARTICLE{kimm2009,
   author = {{Kimm}, T. and {Somerville}, R.~S. and {Yi}, S.~K. and {van den Bosch}, F.~C. and
	{Salim}, S. and {Fontanot}, F. and {Monaco}, P. and {Mo}, H. and
	{Pasquali}, A. and {Rich}, R.~M. and {Yang}, X.},
    title = "{The correlation of star formation quenching with internal galaxy properties and environment}",
  journal = {MNRAS},
archivePrefix = "arXiv",
   eprint = {0810.2794},
     year = 2009,
    month = apr,
   volume = 394,
    pages = {1131-1147},
      doi = {10.1111/j.1365-2966.2009.14414.x},
   adsurl = {http://adsabs.harvard.edu/abs/2009MNRAS.394.1131K}
}

@ARTICLE{koekemoer2011,
       author = {{Koekemoer}, Anton M. and {Faber}, S.~M. and {Ferguson}, Henry C. and {Grogin}, Norman A. and {Kocevski}, Dale D. and {Koo}, David C. and {Lai}, Kamson and {Lotz}, Jennifer M. and {Lucas}, Ray A. and {McGrath}, Elizabeth J. and {Ogaz}, Sara and {Rajan}, Abhijith and {Riess}, Adam G. and {Rodney}, Steve A. and {Strolger}, Louis and {Casertano}, Stefano and {Castellano}, Marco and {Dahlen}, Tomas and {Dickinson}, Mark and {Dolch}, Timothy and {Fontana}, Adriano and {Giavalisco}, Mauro and {Grazian}, Andrea and {Guo}, Yicheng and {Hathi}, Nimish P. and {Huang}, Kuang-Han and {van der Wel}, Arjen and {Yan}, Hao-Jing and {Acquaviva}, Viviana and {Alexander}, David M. and {Almaini}, Omar and {Ashby}, Matthew L.~N. and {Barden}, Marco and {Bell}, Eric F. and {Bournaud}, Fr{\'e}d{\'e}ric and {Brown}, Thomas M. and {Caputi}, Karina I. and {Cassata}, Paolo and {Challis}, Peter J. and {Chary}, Ranga-Ram and {Cheung}, Edmond and {Cirasuolo}, Michele and {Conselice}, Christopher J. and {Roshan Cooray}, Asantha and {Croton}, Darren J. and {Daddi}, Emanuele and {Dav{\'e}}, Romeel and {de Mello}, Duilia F. and {de Ravel}, Loic and {Dekel}, Avishai and {Donley}, Jennifer L. and {Dunlop}, James S. and {Dutton}, Aaron A. and {Elbaz}, David and {Fazio}, Giovanni G. and {Filippenko}, Alexei V. and {Finkelstein}, Steven L. and {Frazer}, Chris and {Gardner}, Jonathan P. and {Garnavich}, Peter M. and {Gawiser}, Eric and {Gruetzbauch}, Ruth and {Hartley}, Will G. and {H{\"a}ussler}, Boris and {Herrington}, Jessica and {Hopkins}, Philip F. and {Huang}, Jia-Sheng and {Jha}, Saurabh W. and {Johnson}, Andrew and {Kartaltepe}, Jeyhan S. and {Khostovan}, Ali A. and {Kirshner}, Robert P. and {Lani}, Caterina and {Lee}, Kyoung-Soo and {Li}, Weidong and {Madau}, Piero and {McCarthy}, Patrick J. and {McIntosh}, Daniel H. and {McLure}, Ross J. and {McPartland}, Conor and {Mobasher}, Bahram and {Moreira}, Heidi and {Mortlock}, Alice and {Moustakas}, Leonidas A. and {Mozena}, Mark and {Nandra}, Kirpal and {Newman}, Jeffrey A. and {Nielsen}, Jennifer L. and {Niemi}, Sami and {Noeske}, Kai G. and {Papovich}, Casey J. and {Pentericci}, Laura and {Pope}, Alexandra and {Primack}, Joel R. and {Ravindranath}, Swara and {Reddy}, Naveen A. and {Renzini}, Alvio and {Rix}, Hans-Walter and {Robaina}, Aday R. and {Rosario}, David J. and {Rosati}, Piero and {Salimbeni}, Sara and {Scarlata}, Claudia and {Siana}, Brian and {Simard}, Luc and {Smidt}, Joseph and {Snyder}, Diana and {Somerville}, Rachel S. and {Spinrad}, Hyron and {Straughn}, Amber N. and {Telford}, Olivia and {Teplitz}, Harry I. and {Trump}, Jonathan R. and {Vargas}, Carlos and {Villforth}, Carolin and {Wagner}, Cory R. and {Wandro}, Pat and {Wechsler}, Risa H. and {Weiner}, Benjamin J. and {Wiklind}, Tommy and {Wild}, Vivienne and {Wilson}, Grant and {Wuyts}, Stijn and {Yun}, Min S.},
        title = "{CANDELS: The Cosmic Assembly Near-infrared Deep Extragalactic Legacy Survey{\textemdash}The Hubble Space Telescope Observations, Imaging Data Products, and Mosaics}",
      journal = {\apjs},
         year = 2011,
        month = dec,
       volume = {197},
       number = {2},
          eid = {36},
        pages = {36},
          doi = {10.1088/0067-0049/197/2/36},
archivePrefix = {arXiv},
       eprint = {1105.3754},
 primaryClass = {astro-ph.CO},
       adsurl = {https://ui.adsabs.harvard.edu/abs/2011ApJS..197...36K}
}

@ARTICLE{kodama1997,
       author = {{Kodama}, T. and {Arimoto}, N.},
        title = "{Origin of the colour-magnitude relation of elliptical galaxies.}",
      journal = {\aap},
         year = 1997,
        month = apr,
       volume = {320},
        pages = {41-53},
          doi = {10.48550/arXiv.astro-ph/9609160},
archivePrefix = {arXiv},
       eprint = {astro-ph/9609160},
 primaryClass = {astro-ph},
       adsurl = {https://ui.adsabs.harvard.edu/abs/1997A&A...320...41K}
}

@ARTICLE{kormendy2009,
       author = {{Kormendy}, John and {Fisher}, David B. and {Cornell}, Mark E. and {Bender}, Ralf},
        title = "{Structure and Formation of Elliptical and Spheroidal Galaxies}",
      journal = {\apjs},
         year = 2009,
        month = may,
       volume = {182},
       number = {1},
        pages = {216-309},
          doi = {10.1088/0067-0049/182/1/216},
archivePrefix = {arXiv},
       eprint = {0810.1681},
 primaryClass = {astro-ph},
       adsurl = {https://ui.adsabs.harvard.edu/abs/2009ApJS..182..216K}
}

@ARTICLE{lange2023,
       author = {{Lange}, Johannes U.},
        title = "{NAUTILUS: boosting Bayesian importance nested sampling with deep learning}",
      journal = {\mnras},
         year = 2023,
        month = oct,
       volume = {525},
       number = {2},
        pages = {3181-3194},
          doi = {10.1093/mnras/stad2441},
archivePrefix = {arXiv},
       eprint = {2306.16923},
 primaryClass = {astro-ph.IM},
       adsurl = {https://ui.adsabs.harvard.edu/abs/2023MNRAS.525.3181L}
}

@ARTICLE{larson1980,
   author = {{Larson}, R.~B. and {Tinsley}, B.~M. and {Caldwell}, C.~N.},
    title = "{The evolution of disk galaxies and the origin of S0 galaxies}",
  journal = {ApJ},
     year = 1980,
    month = may,
   volume = 237,
    pages = {692-707},
      doi = {10.1086/157917},
   adsurl = {http://adsabs.harvard.edu/abs/1980ApJ...237..692L}
}

@ARTICLE{lee2022_gmos_tails,
       author = {{Lee}, Jeong Hwan and {Lee}, Myung Gyoon and {Mun}, Jae Yeon and {Cho}, Brian S. and {Kang}, Jisu},
        title = "{A GMOS/IFU Study of Jellyfish Galaxies in Massive Clusters}",
      journal = {arXiv e-prints},
         year = 2022,
        month = sep,
          eid = {arXiv:2209.07189},
        pages = {arXiv:2209.07189},
archivePrefix = {arXiv},
       eprint = {2209.07189},
 primaryClass = {astro-ph.GA},
       adsurl = {https://ui.adsabs.harvard.edu/abs/2022arXiv220907189L}
}

@ARTICLE{lin2004,
       author = {{Lin}, Yen-Ting and {Mohr}, Joseph J.},
        title = "{K-band Properties of Galaxy Clusters and Groups: Brightest Cluster Galaxies and Intracluster Light}",
      journal = {\apj},
         year = 2004,
        month = dec,
       volume = {617},
       number = {2},
        pages = {879-895},
          doi = {10.1086/425412},
archivePrefix = {arXiv},
       eprint = {astro-ph/0408557},
 primaryClass = {astro-ph},
       adsurl = {https://ui.adsabs.harvard.edu/abs/2004ApJ...617..879L}
}

@ARTICLE{lora2024,
       author = {{Lora}, V. and {Smith}, R. and {Fritz}, J. and {Pasquali}, A. and {Raga}, A.~C.},
        title = "{Dark-matter-free Dwarf Galaxy Formation at the Tips of the Tentacles of Jellyfish Galaxies}",
      journal = {\apj},
         year = 2024,
        month = jul,
       volume = {969},
       number = {1},
          eid = {24},
        pages = {24},
          doi = {10.3847/1538-4357/ad3cda},
archivePrefix = {arXiv},
       eprint = {2404.05676},
 primaryClass = {astro-ph.GA},
       adsurl = {https://ui.adsabs.harvard.edu/abs/2024ApJ...969...24L}
}

@article{luo2023,
   title={Tracing the kinematics of the whole ram-pressure-stripped tails in ESO 137-001},
   volume={521},
   ISSN={1365-2966},
   url={http://dx.doi.org/10.1093/mnras/stad1003},
   DOI={10.1093/mnras/stad1003},
   number={4},
   journal={Monthly Notices of the Royal Astronomical Society},
   publisher={Oxford University Press (OUP)},
   author={Luo, Rongxin and Sun, Ming and Jáchym, Pavel and Waldron, Will and Fossati, Matteo and Fumagalli, Michele and Boselli, Alessandro and Combes, Francoise and Kenney, Jeffrey D P and Li, Yuan and Gronke, Max},
   year={2023},
   month=mar, pages={6266–6283} }

@ARTICLE{manuwal2023,
       author = {{Manuwal}, Aditya and {Stevens}, Adam R.~H.},
        title = "{The relationship between cluster environment and molecular gas content of star-forming galaxies in the EAGLE simulation}",
      journal = {\mnras},
         year = 2023,
        month = aug,
       volume = {523},
       number = {2},
        pages = {2738-2758},
          doi = {10.1093/mnras/stad1587},
archivePrefix = {arXiv},
       eprint = {2212.12187},
 primaryClass = {astro-ph.GA},
       adsurl = {https://ui.adsabs.harvard.edu/abs/2023MNRAS.523.2738M}
}

@ARTICLE{mcconnachie2007,
       author = {{McConnachie}, Alan W. and {Venn}, Kim A. and {Irwin}, Mike J. and {Young}, Lisa M. and {Geehan}, Jonathan J.},
        title = "{Ram Pressure Stripping of an Isolated Local Group Dwarf Galaxy: Evidence for an Intragroup Medium}",
      journal = {ApJL},
         year = 2007,
        month = dec,
       volume = {671},
       number = {1},
        pages = {L33-L36},
          doi = {10.1086/524887},
archivePrefix = {arXiv},
       eprint = {0710.4582},
 primaryClass = {astro-ph},
       adsurl = {https://ui.adsabs.harvard.edu/abs/2007ApJ...671L..33M}
}

@ARTICLE{mcdonald2017,
   author = {{McDonald}, M. and {Allen}, S.~W. and {Bayliss}, M. and {Benson}, B.~A. and
	{Bleem}, L.~E. and {Brodwin}, M. and {Bulbul}, E. and {Carlstrom}, J.~E. and
	{Forman}, W.~R. and {Hlavacek-Larrondo}, J. and {Garmire}, G.~P. and
	{Gaspari}, M. and {Gladders}, M.~D. and {Mantz}, A.~B. and {Murray}, S.~S.
	},
    title = "{The Remarkable Similarity of Massive Galaxy Clusters from z {\tilde} 0 to z {\tilde} 1.9}",
  journal = {ApJ},
archivePrefix = "arXiv",
   eprint = {1702.05094},
     year = 2017,
    month = jul,
   volume = 843,
      eid = {28},
    pages = {28},
      doi = {10.3847/1538-4357/aa7740},
   adsurl = {http://adsabs.harvard.edu/abs/2017ApJ...843...28M}
}

@ARTICLE{maier2019,
       author = {{Maier}, C. and {Hayashi}, M. and {Ziegler}, B.~L. and {Kodama}, T.},
        title = "{Cluster induced quenching of galaxies in the massive cluster XMMXCS J2215.9-1738 at $z \sim 1.5$ traced by enhanced metallicities inside half R$_{200}$}",
      journal = {A\&A},
         year = 2019,
        month = jun,
       volume = {626},
          eid = {A14},
        pages = {A14},
          doi = {10.1051/0004-6361/201935522},
archivePrefix = {arXiv},
       eprint = {1903.09591},
 primaryClass = {astro-ph.GA},
       adsurl = {https://ui.adsabs.harvard.edu/abs/2019A&A...626A..14M}
}

@ARTICLE{mancone2010,
       author = {{Mancone}, Conor L. and {Gonzalez}, Anthony H. and {Brodwin}, Mark and {Stanford}, Spencer A. and {Eisenhardt}, Peter R.~M. and {Stern}, Daniel and {Jones}, Christine},
        title = "{The Formation of Massive Cluster Galaxies}",
      journal = {\apj},
         year = 2010,
        month = sep,
       volume = {720},
       number = {1},
        pages = {284-298},
          doi = {10.1088/0004-637X/720/1/284},
archivePrefix = {arXiv},
       eprint = {1007.1454},
 primaryClass = {astro-ph.CO},
       adsurl = {https://ui.adsabs.harvard.edu/abs/2010ApJ...720..284M}
}

@ARTICLE{marleau2025,
       author = {{Marleau}, F.~R. and {Cuillandre}, J. -C. and {Cantiello}, M. and {Carollo}, D. and {Duc}, P. -A. and {Habas}, R. and {Hunt}, L.~K. and {Jablonka}, P. and {Mirabile}, M. and {Mondelin}, M. and {Poulain}, M. and {Saifollahi}, T. and {S{\'a}nchez-Janssen}, R. and {Sola}, E. and {Urbano}, M. and {Z{\"o}ller}, R. and {Bolzonella}, M. and {Lan{\c{c}}on}, A. and {Laureijs}, R. and {Marchal}, O. and {Schirmer}, M. and {Stone}, C. and {Boselli}, A. and {Ferr{\'e}-Mateu}, A. and {Hatch}, N.~A. and {Kluge}, M. and {Montes}, M. and {Sorce}, J.~G. and {Tortora}, C. and {Venhola}, A. and {Golden-Marx}, J.~B. and {Aghanim}, N. and {Amara}, A. and {Andreon}, S. and {Auricchio}, N. and {Baccigalupi}, C. and {Baldi}, M. and {Balestra}, A. and {Bardelli}, S. and {Battaglia}, P. and {Bender}, R. and {Bodendorf}, C. and {Branchini}, E. and {Brescia}, M. and {Brinchmann}, J. and {Camera}, S. and {Candini}, G.~P. and {Capobianco}, V. and {Carbone}, C. and {Carretero}, J. and {Casas}, S. and {Castellano}, M. and {Cavuoti}, S. and {Cimatti}, A. and {Congedo}, G. and {Conselice}, C.~J. and {Conversi}, L. and {Copin}, Y. and {Courbin}, F. and {Courtois}, H.~M. and {Cropper}, M. and {Da Silva}, A. and {Degaudenzi}, H. and {De Lucia}, G. and {Di Giorgio}, A.~M. and {Dinis}, J. and {Douspis}, M. and {Duncan}, C.~A.~J. and {Dupac}, X. and {Dusini}, S. and {Ealet}, A. and {Farina}, M. and {Farrens}, S. and {Ferriol}, S. and {Fosalba}, P. and {Fotopoulou}, S. and {Frailis}, M. and {Franceschi}, E. and {Fumana}, M. and {Galeotta}, S. and {Garilli}, B. and {George}, K. and {Gillard}, W. and {Gillis}, B. and {Giocoli}, C. and {G{\'o}mez-Alvarez}, P. and {Grazian}, A. and {Grupp}, F. and {Guzzo}, L. and {Hailey}, M. and {Haugan}, S.~V.~H. and {Hoar}, J. and {Hoekstra}, H. and {Holmes}, W. and {Hook}, I. and {Hormuth}, F. and {Hornstrup}, A. and {Hu}, D. and {Hudelot}, P. and {Jahnke}, K. and {Jhabvala}, M. and {Keih{\"a}nen}, E. and {Kermiche}, S. and {Kiessling}, A. and {Kitching}, T. and {Kohley}, R. and {Kubik}, B. and {Kuijken}, K. and {K{\"u}mmel}, M. and {Kunz}, M. and {Kurki-Suonio}, H. and {Lahav}, O. and {Le Mignant}, D. and {Ligori}, S. and {Lilje}, P.~B. and {Lindholm}, V. and {Lloro}, I. and {Maino}, D. and {Maiorano}, E. and {Mansutti}, O. and {Marggraf}, O. and {Markovic}, K. and {Martinet}, N. and {Marulli}, F. and {Massey}, R. and {Maurogordato}, S. and {McCracken}, H.~J. and {Medinaceli}, E. and {Mei}, S. and {Mellier}, Y. and {Meneghetti}, M. and {Merlin}, E. and {Meylan}, G. and {Moresco}, M. and {Moscardini}, L. and {Munari}, E. and {Nakajima}, R. and {Nichol}, R.~C. and {Niemi}, S. -M. and {Padilla}, C. and {Paltani}, S. and {Pasian}, F. and {Pedersen}, K. and {Percival}, W.~J. and {Pettorino}, V. and {Pires}, S. and {Polenta}, G. and {Poncet}, M. and {Popa}, L.~A. and {Pozzetti}, L. and {Raison}, F. and {Rebolo}, R. and {Refregier}, A. and {Renzi}, A. and {Rhodes}, J. and {Riccio}, G. and {Rix}, H. -W. and {Romelli}, E. and {Roncarelli}, M. and {Rossetti}, E. and {Saglia}, R. and {Sapone}, D. and {Scaramella}, R. and {Schneider}, P. and {Secroun}, A. and {Seidel}, G. and {Seiffert}, M. and {Serrano}, S. and {Sirignano}, C. and {Sirri}, G. and {Stanco}, L. and {Tallada-Cresp{\'\i}}, P. and {Taylor}, A.~N. and {Teplitz}, H.~I. and {Tereno}, I. and {Toledo-Moreo}, R. and {Tsyganov}, A. and {Tutusaus}, I. and {Valentijn}, E.~A. and {Valenziano}, L. and {Vassallo}, T. and {Verdoes Kleijn}, G. and {Veropalumbo}, A. and {Wang}, Y. and {Weller}, J. and {Williams}, O.~R. and {Zamorani}, G. and {Zucca}, E. and {Biviano}, A. and {Burigana}, C. and {Scottez}, V. and {Viel}, M. and {Simon}, P. and {Mora}, A. and {Mart{\'\i}n-Fleitas}, J. and {Scott}, D.},
        title = "{Euclid: Early Release Observations {\textendash} Dwarf galaxies in the Perseus galaxy cluster}",
      journal = {\aap},
         year = 2025,
        month = may,
       volume = {697},
          eid = {A12},
        pages = {A12},
          doi = {10.1051/0004-6361/202450799},
archivePrefix = {arXiv},
       eprint = {2405.13502},
 primaryClass = {astro-ph.GA},
       adsurl = {https://ui.adsabs.harvard.edu/abs/2025A&A...697A..12M}
}

@ARTICLE{mayer2006,
   author = {{Mayer}, L. and {Mastropietro}, C. and {Wadsley}, J. and {Stadel}, J. and
	{Moore}, B.},
    title = "{Simultaneous ram pressure and tidal stripping; how dwarf spheroidals lost their gas}",
  journal = {MNRAS},
   eprint = {astro-ph/0504277},
     year = 2006,
    month = jul,
   volume = 369,
    pages = {1021-1038},
      doi = {10.1111/j.1365-2966.2006.10403.x},
   adsurl = {http://adsabs.harvard.edu/abs/2006MNRAS.369.1021M}
}

@ARTICLE{merrifield1994,
       author = {{Merrifield}, Michael R. and {Kuijken}, Konrad},
        title = "{Counterrotating Stars in the Disk of the SAB Galaxy NGC 7217}",
      journal = {\apj},
         year = 1994,
        month = sep,
       volume = {432},
        pages = {575},
          doi = {10.1086/174596},
       adsurl = {https://ui.adsabs.harvard.edu/abs/1994ApJ...432..575M}
}
\bibliographystyle{aasjournal}



\appendix
\section{Tail source size estimate} \label{app:src_size}

\FloatBarrier

\begin{figure}
    \centering
    \includegraphics[width=0.5\textwidth]{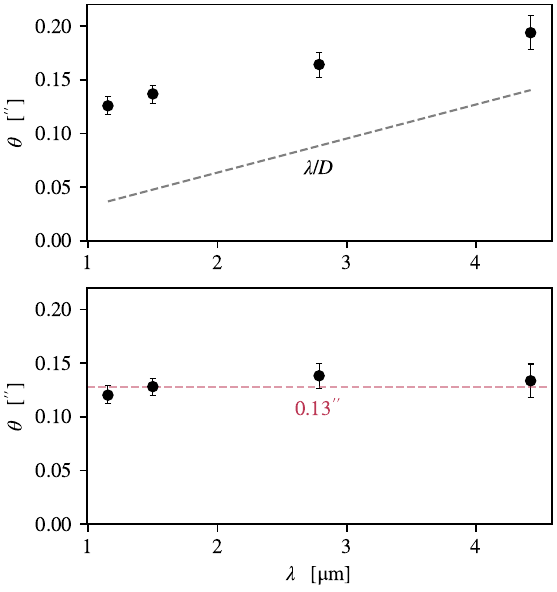}
    \caption{Size estimates for the extra-planar JWST sources in tail region 1. \textit{Top:} Best-fit Gaussian FWHM size averaged over the two JWST sources in tail region 1. Data points correspond to the four JWST filters. For reference we show the diffraction-limited PSF FWHM as a function of wavelength (dashed line), indicating that these sources are only marginally resolved. \textit{Bottom:} Deconvolved FWHM source sizes as a function of wavelength (see text for details).}
    \label{fig:srcsize}
\end{figure}

As an illustrative example, we will consider the source(s) in region 1 (see Fig.~\ref{fig:panels}). This region clearly contains two sources resolved by the JWST images.  To constrain source sizes (and determine whether or not these are simply point sources) we make a small cutout around the sources in region 1 and fit the flux map with the sum of two 2D Gaussian distributions. Fig.~\ref{fig:srcsize}(top) shows the average FWHM source size\footnote{The best-fit FWHMs for the two sources are consistent with each other for all filters, and therefore simplicity we only show the average FWHM.} for each of the four JWST filters.  For reference we also plot $\lambda / D$ corresponding to the diffraction limited PSF size. The sources are marginally resolved with FWHM sizes that are ${\sim}1.5\text{--}3\times$ larger than the diffraction limited PSF. In Fig.~\ref{fig:srcsize}(bottom) we plot `deconvolved sizes' as a function of wavelength, where the deconvolved size is given by
\begin{equation}
    \theta_\mathrm{deconv} = \sqrt{\theta^2 - \theta_\mathrm{diff}^2},
\end{equation}
\noindent
where $\theta$ is the best-fit source FWHM and $\theta_\mathrm{diff}$ is the diffraction limited FWHM. After `deconvolution' all filters agree on a FWHM source size of $0.13\arcsec \approx 1\,\mathrm{kpc}$ corresponding to an effective radius of ${\sim}500\,\mathrm{pc}$. The deconvolution is only formally valid if the underlying source distribution is well described by a symmetric 2D Gaussian (i.e.\ S{\'e}rsic index of 0.5), but still provides an approximate estimate for the characteristic underlying source size.




\end{document}